\documentclass[10pt,twocolumn,twoside]{IEEEtran}
\usepackage{graphicx}
\usepackage[cmex10]{amsmath}
\usepackage{amsfonts}
\usepackage{amssymb}
\usepackage{subfigure}
\usepackage[]{algorithm}
\usepackage{algorithmic}
\usepackage{cite}
\ifCLASSINFOpdf
\else
\fi
\title{Solving Sparse Linear Inverse Problems in Communication Systems: A Deep Learning Approach With Adaptive Depth}
\author{
Wei~Chen,~\IEEEmembership{Senior Member,~IEEE},~Bowen~Zhang,~Shi~Jin,~\IEEEmembership{Senior Member,~IEEE},~Bo~Ai,~\IEEEmembership{Senior Member,~IEEE},~Zhangdui~Zhong,~\IEEEmembership{Senior Member,~IEEE}

\thanks{Wei Chen, Bowen Zhang, Bo Ai and Zhangdui Zhong are with the State Key Laboratory of Rail Traffic Control and Safety, Beijing Jiaotong University, Beijing, China (e-mail: weich,18120171,boai,zhdzhong@bjtu.edu.cn).}
\thanks{Shi Jin is with National Mobile Communications Research Laboratory, Southeast University, Nanjing 210096, P. R. China (e-mail: jinshi@seu.edu.cn).}
\thanks{Corresponding author: Wei Chen}
}

\begin{document}
%
\maketitle
\begin{abstract}
Sparse signal recovery problems from noisy linear measurements appear in many areas of wireless communications. In recent years, deep learning (DL) based approaches have attracted interests of researchers to solve the sparse linear inverse problem by unfolding iterative algorithms as neural networks. Typically, research concerning DL assume a fixed number of network layers. However, it ignores a key character in traditional iterative algorithms, where the number of iterations required for convergence changes with varying sparsity levels. By investigating on the projected gradient descent, we unveil the drawbacks of the existing DL methods with fixed depth. Then we propose an end-to-end trainable DL architecture, which involves an extra halting score at each layer. Therefore, the proposed method learns how many layers to execute to emit an output, and the network depth is dynamically adjusted for each task in the inference phase. We conduct experiments using both synthetic data and applications including random access in massive MTC and massive MIMO channel estimation, and the results demonstrate the improved efficiency for the proposed approach.
\end{abstract}
%
%
\section{Introduction}
In recent years, the sparse linear inverse problem has attracted enormous attention in various applications in wireless communications~\cite{8350399}. For example, in wireless sensor networks, sensors' readings usually have a sparse representation owing to the temporal and spatial correlations. Applying compressive sensing with the sparse linear inverse problem leads to improve energy efficiency~\cite{7390294,7293676}. It has also been applied for detecting active users in massive machine-type communications (MTCs)~\cite{bockelmann2013compressive}, reducing the overhead for channel estimation and feedback in massive multiple-input multiple-output (MIMO) systems~\cite{7913633,8752012}, and enabling sub-Nyquist wideband spectrum sensing in cognitive radio networks~\cite{7559788}.

\par
Mathematically, the ill-posed linear inverse problem can be described as the following optimization problem
\begin{equation}\label{eq:sparse_problem}
\begin{split}
\min_{\mathbf{x}} \quad &\|\mathbf{y}-\mathbf{Ax}\|_2^2\\
\text{s.t.} \quad &f(\mathbf{x})\leq R,
\end{split}
\end{equation}
where $\mathbf{y}\in\mathbb{R}^n$ is a vector of measurements, $\mathbf{A}\in \mathbb{R}^{n\times m}$ is the measurement matrix, $\mathbf{x}\in\mathbb{R}^{m}$ is the unknown sparse signal, $f(\mathbf{x})$ is a function that enforces sparsity, and $R$ is a parameter that is either pre-determined or tuned. The most straightforward choice of $f$ is the $\ell_0$ pseudo-norm that counts the number of nonzero elements in a vector. However, (\ref{eq:sparse_problem}) with the $\ell_0$ pseudo-norm is an intractable combinatorial optimization problem. Therefore, instead of using the $\ell_0$ pseudo-norm, efficient approximations are developed in literature. Popular techniques include convex relaxations such as $\ell_1$ norm minimization, and greedy approaches such as orthogonal matching pursuit (OMP) and iterative hard-thresholding (IHT). For the case of noiseless measurements, these algorithms are capable of finding the maximally sparse signal with a high probability in restricted regimes, although they have varying degrees of computational complexity. For the case that measurements are corrupted by noise, these algorithms conduct many iterations until some stopping criterion is achieved.

\par
Pursuing high reconstruction accuracy is not the only goal in real world. In some applications, we are requested to solve the problem with a time constraint. For example, 5th-generation (5G) wireless communications have latency requirements specified in IMT-2020~\cite{wp5d2017guidelines} so that algorithms with high computational complexity are not desired. To this end, the goal of algorithm development becomes minimizing the cost function with a fixed number of operations that can be performed to recover the sparse vector. A promising direction of growing interests nowadays is to employ deep learning (DL) techniques to develop fast yet accurate algorithms for the sparse linear inverse problem, where neural networks with fixed depth can be constructed by unfolding traditional iterative algorithms. Unlike traditional iterative algorithms where no parameter is learnable, parameters in neural networks are learned over a set of training data pairs $\{\mathbf{x}_j,\mathbf{y}_j\}$ sampled from some distribution $\mathcal{P}(\mathbf{x},\mathbf{y})$. It is computationally expensive to train the neural networks and learn these parameters, while the training can be done in the off-line manner. Empirical results show that the trained neural networks generalize well to new samples drawn from the same distribution and can successfully recover unseen sparse signals with a significantly reduced averaged number of iterations/layers in the inference~\cite{gregor2010learning,7934066,xin2016maximal,he2017bayesian}.

\par
DL provides highly successful neural networks for various applications in computer vision, image processing, natural language and communications processing~\cite{lecun2015deep,8882268}.
In the context of solving the sparse linear inverse problem, by unfolding each iterative step of a sparse recovery algorithm, we obtain a signal-flow graph whose variables can be learned in a supervised manner via stochastic gradient descent. For example, by exploiting the basic structure of iterative shrinkage-thresholding algorithms (ISTA) whose parameters in updating rules are determinate, the learned ISTA (LISTA)~\cite{gregor2010learning} adjusts these parameters via supervised learning. In comparison to the ISTA, the LISTA uses one to two orders of magnitude less iterations and achieves the same reconstruction accuracy in the inference phase~\cite{gregor2010learning}. In~\cite{liu2019alista}, Liu et al. propose to use analytic parameters and only learn a series of scalars for thresholding and step sizes, which simplifies the training phase. Chen et al. propose LISTA-CPSS to improves the convergence rate of LISTA, which introduces a partial weight coupling structure and support selection to LISTA~\cite{Chen2018theoretical}. Inspired by the approximate message passing (AMP) algorithm and the vector AMP (VAMP) algorithm, Borgerding et al. propose a learned AMP (LAMP) architecture and a learned VAMP (LVAMP) architecture, respectively~\cite{7934066}. More DL based algorithms for the sparse linear inverse problem will be reviewed in the next section. In contrast with the classical optimization approach where an expert designs a heuristic algorithm, DL based approach for solving optimization problem consumes lots of data to characterize the distribution of interest, works in the space of algorithm designs and learns very highly parameterized algorithms to express a combinatorial space of heuristic design choices.

\begin{figure}[!t]
\centering
\subfigure[Mean]{\includegraphics[width=0.45\linewidth]{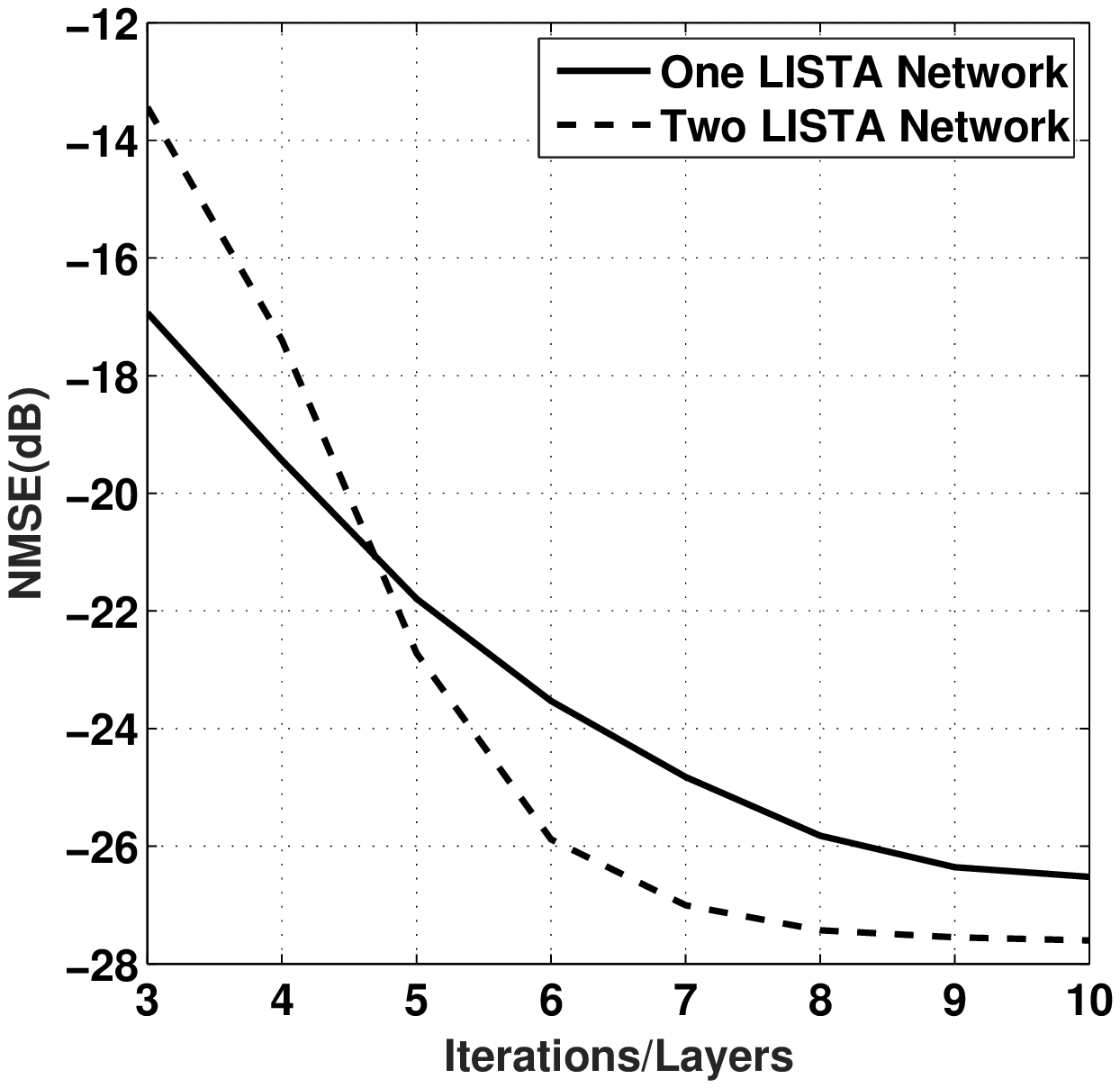}
\label{measurement}}
\subfigure[Standard deviation]{\includegraphics[width=0.45\linewidth]{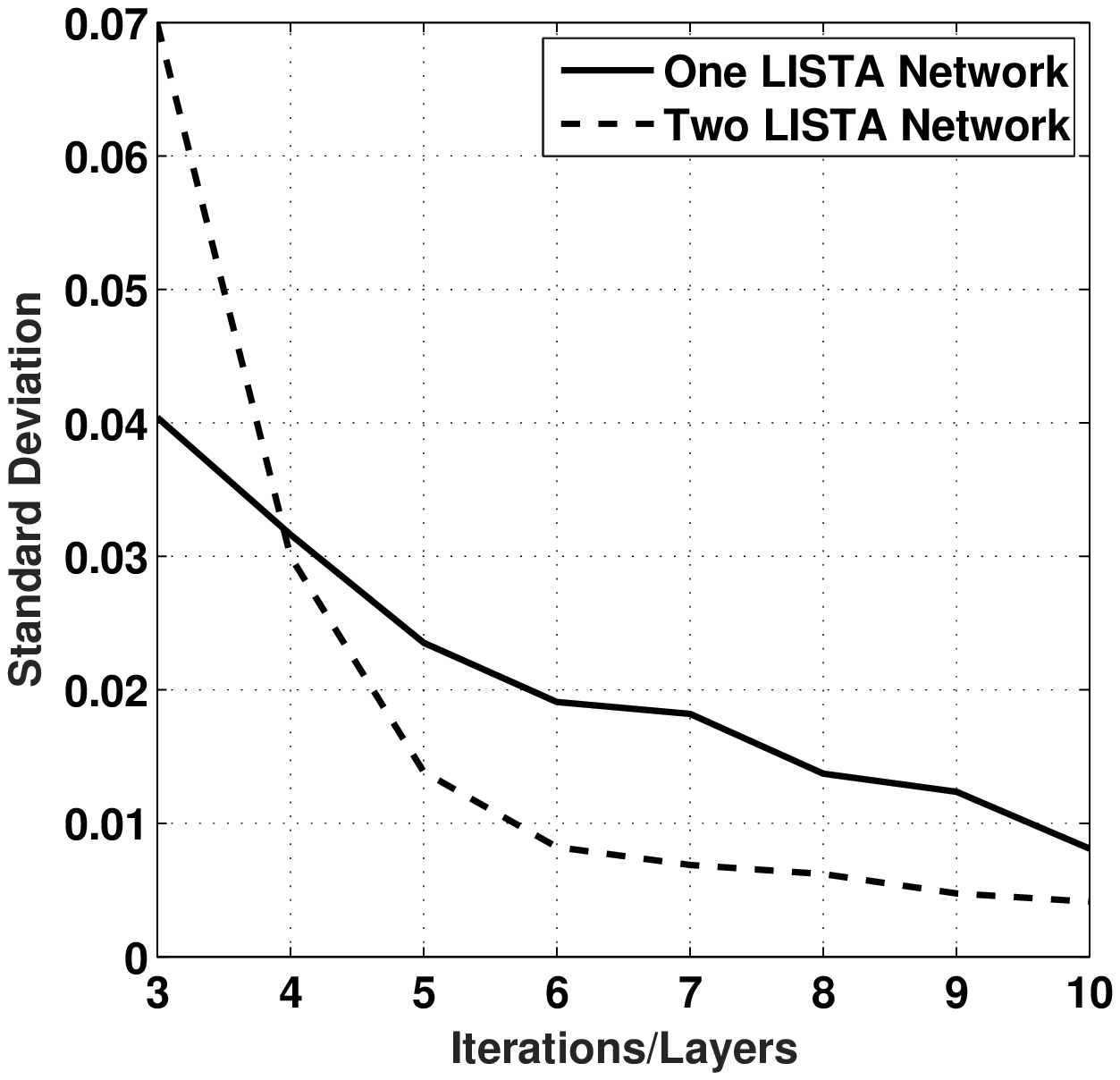}
\label{dimension}}
\caption{Evaluation of reconstruction error (two LISTA networks vs. one LISTA network, $n=40$ and $m=200$).} \label{fig:one_two_LISTA}
\end{figure}

\par
Most of the study on DL based approaches for sparse linear inverse problems focuses on further improving the reconstruction accuracy, and ignores a key difference between the DL based approaches and traditional iterative algorithms. That is, traditional iterative algorithms can automatically adjust the number of iterations for different tasks with varying sparsity levels and/or the noise levels~\cite{8107507}, while the network depth of DL approaches is pre-determined in the training and the computing time in inference is proportional to the network depth. In~\cite{8107507}, Samet Oymak et al. characterize the time-data (including sparsity, number of measurements, etc.) trade-off for optimization problems used for solving linear inverse problems, and their results show that more iterations are needed for problems with a higher sparsity level. Current DL based approaches for sparse linear inverse problems are unable to adjust the network depth.

\par
Fixed network depth leads to two shortcomings: i) the waste of computing resource when the neural network is used for ``easy'' tasks, e.g., recovering a very sparse $\mathbf{x}$; and ii) unsatisfied reconstruction quality when the neural network is used for ``hard'' tasks. An illustrating example is shown in Fig. 1, where a half of the signals for recovery have the sparsity level $s=2$ (seen as ``easy'' tasks) and the other half signals have the sparsity level $s=4$ (seen as ``hard'' tasks). We compare two approaches: i) training one LISTA network of depth $L$ for recovering mixed signals of sparsity $s\in\{2,4\}$; ii) training two LISTA networks independently, i.e., a short LISTA network with depth $L-2$ for recovering signals of sparsity $s=2$ and a long LISTA network with depth $L+2$ for recovering signals of sparsity $s=4$. Signals of different sparsity levels emit at the end of the corresponding network. We change $L$ from 3 to 10, and train and test different one-LISTA and two-LISTA networks for every $L$. With this setting, the averaged number of executed layers is $L$ for both the approaches. According to the study shown in Fig. 1, with the same averaged number of executed layers, using two LISTA networks with different depths attains a better averaged reconstruction quality and also a smaller variance of reconstruction error. However, in practical applications, the sparsity level of each signal is unknown generally, and thus one cannot determine which LISTA network should be used in the inference phase (although one can train multiple networks for different sparsity levels). Furthermore, using two independent LISTA networks clearly uses twice the amount of resources compared to a single LISTA network even though they both use $L$ layers on an average. To address this issue, it calls for a way to make a single network adaptively adjust depth for different tasks.

\par
Adaptive computation time (ACT)~\cite{graves2016adaptive} was recently proposed as a way for recurrent neural networks (RNNs) to do different amounts of computation. Although ACT has shown to be a promising technique in various applications from character prediction~\cite{graves2016adaptive} to image classification~\cite{figurnov2017spatially}. There are several drawbacks that make the ACT architecture inefficient for solving the sparse linear inverse problem in (\ref{eq:sparse_problem}). Firstly, the architecture of the ACT leads to a discontinuous cost function, which is hard to optimize generally; Secondly, the halting score of each layer is computed only from the output of that layer, which fails to exploit the information in $\mathbf{y}$ and $\mathbf{A}$ and thus leads to a bad halting decision so that the reconstruction quality of $\mathbf{x}$ is poor; Thirdly, the output of the network (in both the training and inference) is a weighted sum of the outputs of all layers, which differs with iterative algorithms that emit the output of the last iteration. Lastly, the ACT network needs to be trained again, if one want to vary the averaged depth (in inference) to adapt to different average performance requirements. Recurrent neural networks (RNNs), e.g., the long short-term memory (LSTM), are proposed in literature to solve sparse linear inverse problems~\cite{he2017bayesian}. However, how to determine the number of iterations is not addressed.

\par
The goal of this work is to answer the following two key questions for the sparse linear inverse problem that is raised in communication systems:
\begin{itemize}
  \item How to characterize the benefits brought by DL based methods with adaptive depth?
  \item How to design a neural network architecture for adapting depth to solve sparse linear inverse problems with varying sparsity levels?
\end{itemize}
To answer the first question, we extend the convergence analysis of the projected gradient descent (PGD) method in the cases of using learned gradients and adaptive depth. To answer the other question, we propose an end-to-end trainable architecture that can dynamically adapt the number of executed layers for each coming task. The contributions of our work are as following
\begin{itemize}
  \item	Method: The proposed method is the first work that improves DL methods for solving sparse linear inverse problems by using adaptive depth. The proposed structure can be incorporated into many existing neural networks and applied to various applications in wireless communications.
  \item Analysis: We develop novel theoretical analysis on the convergence of the learned PGD algorithm and the benefits of adaptive depth using the random matrix theory; the established theoretical results help us to analyze our design of halting score and cost function (this part is newly added in the revised paper).
  \item Application/Evaluation: We demonstrate the benefits brought by the proposed method in two applications including random access in massive MTC and massive MIMO channel estimation.
\end{itemize}
The proposed architecture includes several novel designs including: i) learning a linear mapping for the residual of the output in each iteration, which improves the accuracy of the prediction of the reconstruction error and the quality of the halting score, ii) a continuous and differentiable cost function that involves regularizers on halting scores and the weighted summation of the reconstruction error of all layers, where the weights are inverse of the halting scores, and iii) a nonsymmetric training and inference processes, where the computation is dynamically adapted only in the inference. We evaluate the proposed method which achieves improved time-accuracy-hardness trade-off in the experiments. Note that the proposed method enables early exit when testing, while early stopping is a form of regularization used to avoid overfitting when training a learner.

\par
The rest of the paper is organized as follows: Section \uppercase\expandafter{\romannumeral2} describes related work on DL based approaches and the projected gradient descent (PGD) method in traditional iterative algorithms. Section \uppercase\expandafter{\romannumeral3} provides extensions on the PGD
convergence analysis in the cases of using learned gradients and adaptive depth to unveil the benefits brought by DL based
methods with adaptive depth. In \uppercase\expandafter{\romannumeral4}, a novel method is developed to enable neural networks to learn how many layers to execute to emit the output in solving the sparse linear inverse problem. Numerical results are presented in Section \uppercase\expandafter{\romannumeral5}, followed by conclusions in Section \uppercase\expandafter{\romannumeral6}.

\section{Background}
\subsection{Related Work on DL Based Approaches}
DL based approaches are usually designed under the guidance of model-based iterative methods. A popular algorithmic approach to solving the sparse linear inverse problem is the ISTA that considers the convex relaxed optimization problem
\begin{equation}\label{eq:L1_problem}
\begin{split}
\min_{\mathbf{x}} \quad &\frac{1}{2}\|\mathbf{y}-\mathbf{Ax}\|_2^2 + \lambda\|\mathbf{x}\|_1,
\end{split}
\end{equation}
where $\lambda>0$ is a tunable parameter. The ISTA first computes the gradient of the quadratic objective at $\mathbf{x}_t$
\begin{equation}\label{eq:ISTA_gradient}
\frac{\partial \frac{1}{2}\|\mathbf{y}-\mathbf{Ax}\|_2^2}{\partial \mathbf{x}}\bigg|_{\mathbf{x}=\mathbf{x}_t} = \mathbf{A}^T\mathbf{Ax}_t-\mathbf{A}^T\mathbf{y},
\end{equation}
and then conducts the unconstrained gradient step
\begin{equation}\label{eq:ISTA_unconstrained}
\mathbf{z}=\mathbf{x}_{t}-\beta\big(\mathbf{A}^T\mathbf{Ax}_t-\mathbf{A}^T\mathbf{y}\big),
\end{equation}
where $\beta$ is the step size. Lastly, the ISTA applies a proximal mapping $\mathcal{S}_{\lambda}(\cdot)$
\begin{equation}\label{eq:ISTA_mapping}
\mathcal{S}_{\lambda}(\mathbf{z})=\arg\min_{\mathbf{x}}\frac{1}{2}\|\mathbf{x}-\mathbf{z}\|_2^2 + \lambda\|\mathbf{x}\|_1,
\end{equation}
which is an element-wise soft thresholding shrinkage function and can be written as
\begin{equation}\label{eq:ISTA_soft_thresholding}
[\mathcal{S}_{\lambda}(\mathbf{z})]_i = \text{sgn}(z_i)\max \{|z_i|-\lambda,0\}.
\end{equation}
Therefore, the each iteration of the ISTA can be expressed as
\begin{equation}\label{eq:ISTA_iteration}
\mathbf{x}_{t+1} = \mathcal{S}_{\lambda}\big((\mathbf{I}-\beta\mathbf{A}^T\mathbf{A})\mathbf{x}_{t}+\beta\mathbf{A}^T\mathbf{y}\big).
\end{equation}
The ISTA only involves matrix multiplications and element-wise nonlinear operations, and is guaranteed to converge given $\beta\leq \frac{1}{\|\mathbf{A}\|_2^2}$. However, the convergence of the ISTA is somewhat slow and various modifications of the iteration step (\ref{eq:ISTA_iteration}) have been studied to speed up convergence~\cite{beck2009fast,Zulfiquar2015}.

\begin{figure*}[!tb]%
\centering%
\includegraphics[width=0.8\textwidth]{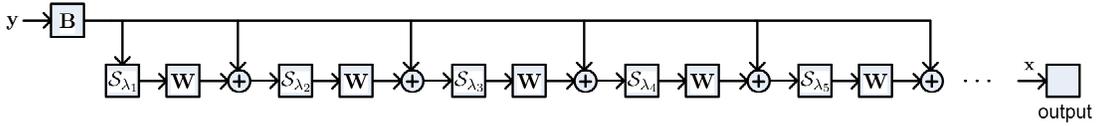}%
\DeclareGraphicsExtensions. \caption{The network structures of LISTA.} \label{fig:LISTA}
\end{figure*}%
\par
The iterations of the ISTA, i.e., (\ref{eq:ISTA_iteration}), can be unfolded into a neural network with fixed weights shared by all layers. In the LISTA~\cite{gregor2010learning}, Gregor and LeCun consider a more general iteration step, given by
\begin{equation}\label{eq:ISTA_iteration2}
\mathbf{x}_{t+1} = \mathcal{S}_{\lambda_t}(\mathbf{Wx}_{t}+\mathbf{By}),
\end{equation}
where $\lambda_t$, $\mathbf{W}\in \mathbb{R}^{m\times m}$ and $\mathbf{B}\in \mathbb{R}^{m\times n}$ are learned from a large set of training data by using stochastic gradient descent. Here, $1\leq t\leq L$ and $L$ denotes the total number of layers. The network structure of the LISTA is shown in Fig.\ref{fig:LISTA}. In comparison to the ISTA, a prominent advantage of the LISTA is the reduction of the number of iterations/layers to attain satisfied accuracy.

\par
Akin to the LISTA, different designs of neural networks have been proposed by ``unfolding'' traditional iterative algorithms ~\cite{7934066,he2017bayesian,sun2016deep,andrychowicz2016learning,hershey2014deep,7442798,xin2016maximal,wang2016learning,ablin2019learning,8695874,7952977}. For example, Borgerding et al. propose neural networks inspired by the AMP algorithms in~\cite{7934066}, He et al. examine the structural similarities between sparse Bayesian learning algorithms and the LSTM networks in~\cite{he2017bayesian}, and Yang et al. unfold the alternating direction method of multipliers (ADMM) algorithm and propose ADMM-net in~\cite{sun2016deep}. Ablin et al. propose a network architecture where only the step sizes of ISTA are learned~\cite{ablin2019learning}. Ito et al. propose a trainable iterative soft thresholding algorithm that includes a linear estimation unit and a minimum mean squared error (MMSE) estimator based shrinkage unit~\cite{8695874}. Wisdom et al. show the benefit of using a stacked RNN with backpropagation using supervised data for sequential sparse recovery~\cite{7952977}. Bai et al. propose to use block restrictive activation nonlinear unit to capture the block sparse structure in~\cite{vtc2019_bai}. In~\cite{8353153}, a learned denoising-based approximate message passing (LDAMP) network is proposed to estimate the sparse channel coefficients for millimeter-wave massive multiple-input and multiple-output (MIMO) systems. In~\cite{8646357}, a learned network designed by unfolding the orthogonal AMP (OAMP) is proposed for MIMO detection. All of these existing approaches consider networks with a pre-determined number of layers.

\subsection{Projected Gradient Descent (PGD)}
The PGD method can be applied to solve the constrained optimization problem in (\ref{eq:sparse_problem}). In each iteration, we move in the direction of the negative gradient, and then project onto the feasible set. Each iteration of the PGD method can be expressed as
\begin{equation}\label{eq:PGD_iteration}
\mathbf{x}_{t+1} = \mathcal{P}_{\mathcal{K}}\big((\mathbf{I}-\beta\mathbf{A}^T\mathbf{A})\mathbf{x}_{t}+\beta\mathbf{A}^T\mathbf{y}\big),
\end{equation}
where $\mathcal{K}$ denotes the set
\begin{equation}\label{eq:PGD_set}
\mathcal{K} = \{\mathbf{x}\in \mathbb{R}^m:\ f(\mathbf{x})\leq R\},
\end{equation}
and $\mathcal{P}_{\mathcal{K}}(\cdot)$ denotes the Euclidean projection onto the set $\mathcal{K}$
\begin{equation}\label{eq:PGD_projection}
\mathcal{P}_{\mathcal{K}}(\mathbf{x}) = \arg\min_{\mathbf{z}\in \mathcal{K}}\ \|\mathbf{x}-\mathbf{z}\|_2^2.
\end{equation}
For the special case that $\mathcal{K}$ is the $\ell_0$-ball, the PGD method becomes the iterative hard thresholding (IHT) method~\cite{blumensath2009iterative}. For the case that $\mathcal{K}$ is the $\ell_1$-ball, $\mathcal{P}_{\mathcal{K}}$ is the same as the proximal mapping (\ref{eq:ISTA_mapping}) in the ISTA except the soft thresholding parameter $\lambda$ varies depending on the projected vector $\mathbf{x}$.

\par
The PGD method is proved to converge to the true signal in the noiseless case~\cite{8107507}, when the number of measurements, $m$, is sufficient and the parameter $R$ is tuned perfectly to $R=f(\mathbf{x})$. In practice $f(\mathbf{x})$ might be unknown so that $R\neq f(\mathbf{x})$. Theorem~\ref{thm:stability_PGD} below provides the convergence rate and stability analysis for the PGD with perfect and imperfect tuning parameter.

\par
Before introducing the result, the following definition is needed.

\newtheorem{definition}{Definition}
\begin{definition}[Descent Set and Cone]
The set of descent of the function $f$ at a point $\mathbf{x}$ is defined as
\begin{equation}\label{eq:Descent_Set_Set}
\mathcal{D}_{f}(\mathbf{x}) = \big\{\mathbf{d}\in \mathbb{R}^m:f(\mathbf{x+d})\leq f(\mathbf{x})\big\}.
\end{equation}
The tangent cone $\mathcal{C}_f(\mathbf{x})$ is the conic hull of the descent set $\mathcal{D}_{f}(\mathbf{x})$. That is, the smallest closed cone $\mathcal{C}_f(\mathbf{x})$ obeying $\mathcal{D}_{f}(\mathbf{x})\subseteq \mathcal{C}_f(\mathbf{x})$.
\end{definition}

\begin{definition}[Gaussian Mean Width]
The Gaussian mean width of a set $\mathcal{C}\in \mathbb{R}^m$ is defined as
\begin{equation}\label{eq:Gaussian_Width}
\omega(\mathcal{C})= \mathbb{E}_\mathbf{g}[\sup\limits_{\mathbf{z}\in \mathcal{C}\cap\mathcal{B}^m}\langle\mathbf{g},\mathbf{z}\rangle],
\end{equation}
where $\mathbf{g}\sim \mathcal{N}(\mathbf{0},\mathbf{I})$, and $\mathcal{B}^m$ denotes $m$-dimensional $\ell_2$-ball of radius 1.
\end{definition}

The following definition of phase transition  characterizes the minimum required number of measurements for successful reconstruction.
\begin{definition}[Phase Transition Function]
Let $\mathbf{x}\in \mathbb{R}^m$ be an arbitrary vector, $f:\mathbb{R}^m\rightarrow \mathbb{R}$ be a proper function, $\mathcal{C}_f(\mathbf{x})$ be the tangent cone of $f$ at $\mathbf{x}$, $\omega=\omega(\mathcal{C}_f(\mathbf{x}))$ and $\phi(t)=\sqrt{2}\frac{\Gamma(\frac{t+1}{2})}{\Gamma(\frac{t}{2})}\approx\sqrt{t}$. The phase transition function is defined as
\begin{equation}\label{eq:Phase_Transition}
n_0(\mathbf{x},\mathcal{C}_f,\eta) = \phi^{-1}(\omega+\eta) \approx (\omega+\eta)^2,
\end{equation}
where $\eta$ is a parameter controlling the probability of success.
\end{definition}

We now introduce the convergence rate and stability analysis provided in~\cite{8107507} for the PGD in both the case of perfect tuning parameter and the case of imperfect tuning parameter.
\newtheorem{theorem}{Theorem}
\begin{theorem}[Theorem 5 and Theorem 9 in~\cite{8107507}]\label{thm:stability_PGD}
Let $\mathbf{x}\in \mathbb{R}^m$ be an arbitrary vector, $f:\mathbb{R}^m\rightarrow \mathbb{R}$ be a norm function, $\mathcal{K} = \{\mathbf{x}\in \mathbb{R}^m:\ f(\mathbf{x})\leq R\}$ be a set, $\mathcal{C}=\mathcal{C}_f(\mathbf{x})$ be the tangent cone of $f$ at $\mathbf{x}$, $\mathbf{A}\in \mathbb{R}^{n\times m}$ have independent $\mathcal{N}(\mathbf{0},1)$ entries and $\mathbf{y}=\mathbf{Ax}\in\mathbb{R}^n$. Let $\kappa_f$ be a constant that is equal to 1 for convex function $f$ and equal to 2 for non-convex function $f$. Set the learning parameter to $\beta=\frac{1}{2}\left(\frac{\Gamma(\frac{n}{2})}{\Gamma(\frac{n+1}{2})}\right)^2\approx \frac{1}{n}$. Let $n_0$ defined in (\ref{eq:Phase_Transition}) be the minimum number of measurements required by the phase transition curve, and $\rho=\sqrt{8\kappa_f^2\frac{n_0}{n}}$. Then as long as
\begin{equation}\label{eq:stability_PGD1}
n>8\kappa_f^2 n_0,
\end{equation}
and starting from the initial point $\mathbf{x}_0=\mathbf{0}$, the update (\ref{eq:PGD_iteration}) obeys the following conditions with probability at least $1-8e^{-\frac{\eta^2}{8}}$,
\begin{equation}\label{eq:stability_PGD2}
\|\mathbf{x}_t-\mathbf{x}\|_2\leq \rho^t\|\mathbf{x}\|_2+\frac{3-\kappa_f}{1-\rho}\|\mathbf{x}-\mathcal{P}_{\mathcal{K}}(\mathbf{x})\|_2
\end{equation}
for all $R<f(\mathbf{x})$,
\begin{equation}\label{eq:stability_PGD3}
\|\mathbf{x}_t-\mathbf{x}\|_2\leq \rho^t\|\mathbf{x}\|_2
\end{equation}
for $R=f(\mathbf{x})$, and
\begin{equation}\label{eq:stability_PGD4}
\|\mathbf{x}_t-\mathbf{x}\|_2\leq \rho^t\|\mathbf{x}\|_2+\frac{\kappa_f+1+2\rho}{1-\rho}\left(\frac{R}{f(\mathbf{x})}-1\right)\|\mathbf{x}\|_2
\end{equation}
for all $R>f(\mathbf{x})$.
\end{theorem}

\par
We would like to emphasize that in comparison to (\ref{eq:stability_PGD3}), i.e., the case with perfect tuning parameter, the extra errors in (\ref{eq:stability_PGD2}) and (\ref{eq:stability_PGD4}) result from imperfect tuning parameter $R$ and go to zeros as $R\rightarrow f(\mathbf{x})$.

\section{Extensions on the PGD Convergence Analysis}
The PGD algorithm can be seen as a neural network with pre-determined parameters, i.e., omitting the training process. In this section, we extend the PGD convergence analysis for two different cases, i.e., PGD with a learned gradient and PGD with adaptive depth, which not only provides the motivations of this work but also shed lights on the proposed method for enhancing DL based neural networks for sparse linear inverse problems.

\subsection{Enhancing the PGD via Learned Gradients}
In the presence of some training data, one could improve the convergence of the PGD algorithm via computing gradients with learned parameters. For example, instead of computing (\ref{eq:ISTA_gradient}), i.e., the gradient of the unconstraint objective in (\ref{eq:sparse_problem}), we consider a learned gradient given by
\begin{equation}\label{eq:Learned_gradient}
g(\mathbf{x}_t,\mathbf{A},\mathbf{y}) = \mathbf{B}\mathbf{Ax}_t-\mathbf{B}\mathbf{y},
\end{equation}
where $\mathbf{B}\in \mathbb{R}^{m\times n}$ contains parameters learned from training data. Then the unconstrained gradient step becomes\footnote{The step size is embedded in the learned matrix $\mathbf{B}$.}
\begin{equation}\label{eq:Learned_gradient_step}
\mathbf{z}=\mathbf{x}_{t}-\big(\mathbf{B}\mathbf{Ax}_t-\mathbf{B}\mathbf{y}\big).
\end{equation}
With this modification, each iteration of the PGD method can be written as
\begin{equation}\label{eq:Learned_PGD_iteration}
\mathbf{x}_{t+1} = \mathcal{P}_{\mathcal{K}}\big((\mathbf{I}-\mathbf{B}\mathbf{A})\mathbf{x}_{t}+\mathbf{By}\big).
\end{equation}

\par
The following theorem establishes the convergence performance of the PGD algorithm in the noise case with a learned $\mathbf{B}$.

\begin{theorem}\label{thm:learned_PGD}
Let $\mathbf{x}\in \mathbb{R}^m$ be an arbitrary vector in the set $\mathcal{K} = \{\mathbf{x}\in \mathbb{R}^m:\ f(\mathbf{x})\leq R\}$, $f:\mathbb{R}^m\rightarrow \mathbb{R}$ be a proper function, $\mathcal{C}=\mathcal{C}_f(\mathbf{x})$ be the tangent cone of $f$ at $\mathbf{x}$, $\mathbf{A}\in \mathbb{R}^{n\times m}$ be a linear mapping, $\boldsymbol{\omega}\in\mathbb{R}^n$ be a noise vector and $\mathbf{y}=\mathbf{Ax}+\boldsymbol{\omega}\in\mathbb{R}^n$. Let $\kappa_f$ be a constant that is equal to 1 for convex function $f$ and equal to 2 for non-convex function $f$. Starting from the initial point $\mathbf{x}_0=\mathbf{0}$, the update (\ref{eq:Learned_PGD_iteration}) obeys
\begin{equation}\label{eq:learned_PGD2}
\|\mathbf{x}_t-\mathbf{x}\|_2\leq (\kappa_f\rho(\mathbf{B}))^t\|\mathbf{x}\|_2
+\kappa_f\frac{1-(\kappa_f\rho(\mathbf{B}))^t}{1-\kappa_f\rho(\mathbf{B})}\xi(\mathbf{B})\|\boldsymbol{\omega}\|_2,
\end{equation}
where
\begin{equation}\label{eq:learned_PGD3}
\rho(\mathbf{B})= \sup\limits_{\mathbf{u},\mathbf{v}\in \mathcal{C}\cap \mathcal{B}^m} \mathbf{u}^T(\mathbf{I}-\mathbf{B}\mathbf{A})\mathbf{v},
\end{equation}
and
\begin{equation}\label{eq:learned_PGD4}
\xi(\mathbf{B})= \sup\limits_{\mathbf{u}\in \mathcal{C}\cap \mathcal{B}^m} {\mathbf{u}}^T\mathbf{B}\frac{\boldsymbol{\omega}}{\|\boldsymbol{\omega}\|_2}.
\end{equation}
\end{theorem}

\begin{IEEEproof}
According to the definition of the tangent cone, we have
\begin{equation}\label{eq:learned_PGD_proof1}
\begin{split}
\mathcal{D}&=\big\{\mathbf{d}:f(\mathbf{x+d})\leq f(\mathbf{x})\big\}\\
&=\big\{\mathbf{z}-\mathbf{x}:f(\mathbf{z})\leq f(\mathbf{x})\big\}\\
&=\big\{\mathbf{z}-\mathbf{x}:\mathbf{z}\in \mathcal{K}\big\}.
\end{split}
\end{equation}
By applying (\ref{eq:learned_PGD_proof1}), the error at the $(t+1)$th iteration can be rewritten as
\begin{equation}\label{eq:learned_PGD_proof2}
\begin{split}
\|\mathbf{x}_{t+1}-\mathbf{x}\|_2&=\|\mathcal{P}_{\mathcal{K}}((\mathbf{I}-\mathbf{B}\mathbf{A})\mathbf{x}_{t}+\mathbf{By})-\mathbf{x}\|_2\\
&=\|\mathcal{P}_{\mathcal{K}}((\mathbf{I}-\mathbf{B}\mathbf{A})\mathbf{x}_{t}+\mathbf{BAx}+\mathbf{B}\boldsymbol{\omega})-\mathbf{x}\|_2\\
&=\|\mathcal{P}_{\mathcal{D}}((\mathbf{I}-\mathbf{B}\mathbf{A})(\mathbf{x}_{t}-\mathbf{x})+\mathbf{B}\boldsymbol{\omega})\|_2.\\
\end{split}
\end{equation}
Lemma 18 in~\cite{8107507} proves $\|\mathcal{P}_{\mathcal{D}}(\mathbf{z})\|_2\leq \kappa_f\|\mathcal{P}_{\mathcal{C}}(\mathbf{z})\|_2$ for any nonempty set $\mathcal{D}$ that contains $0$. Therefore, the error at the $(t+1)$th iteration can be upper bounded by
\begin{equation}\label{eq:learned_PGD_proof3}
\begin{split}
&\|\mathbf{x}_{t+1}-\mathbf{x}\|_2\\
\leq &\kappa_f\|\mathcal{P}_{\mathcal{C}}((\mathbf{I}-\mathbf{B}\mathbf{A})(\mathbf{x}_{t}-\mathbf{x})+\mathbf{B}\boldsymbol{\omega})\|_2\\
\leq &\sup\limits_{\mathbf{u}\in \mathcal{C}\cap \mathcal{B}^m} \kappa_f{\mathbf{u}}^T(\mathbf{I}-\mathbf{B}\mathbf{A})(\mathbf{x}_{t}-\mathbf{x})+\sup\limits_{\mathbf{u}\in \mathcal{C}\cap \mathcal{B}^m} \kappa_f{\mathbf{u}}^T\mathbf{B}\boldsymbol{\omega}\\
\leq &\sup\limits_{\mathbf{u},\mathbf{v}\in \mathcal{C}\cap \mathcal{B}^m} \kappa_f{\mathbf{u}}^T(\mathbf{I}-\mathbf{B}\mathbf{A})\mathbf{v}\|\mathbf{x}_{t}-\mathbf{x}\|_2
+\sup\limits_{\mathbf{u}\in \mathcal{C}\cap \mathcal{B}^m} \kappa_f{\mathbf{u}}^T\mathbf{B}\boldsymbol{\omega}\\
= &\kappa_f\rho(\mathbf{B})\|\mathbf{x}_{t}-\mathbf{x}\|_2+\xi(\mathbf{B})\|\boldsymbol{\omega}\|_2.
\end{split}
\end{equation}
Following the proof of the Theorem 2 in~\cite{8107507}, one can apply the above equation recursively to conclude the proof.
\end{IEEEproof}

\emph{Remark:} The convergence result in (\ref{eq:learned_PGD2}) generalizes the result of Theorem 2 in~\cite{8107507}, where $\mathbf{B}=\mathbf{A}^T$. As $\min\limits_{\mathbf{B}} \rho(\mathbf{B})\leq \rho(\mathbf{A}^T)$ and $\min\limits_{\mathbf{B}} \xi(\mathbf{B})\leq \xi(\mathbf{A}^T)$, it unveils the potential of improving the convergence of the PGD algorithm by using learned parameters. Furthermore, better convergence performance is expected when $\mathbf{I}-\mathbf{B}\mathbf{A}$ in (\ref{eq:Learned_PGD_iteration}) is replaced by a learned matrix $\mathbf{W}$, which gives more flexibility to the algorithm design. Here we consider the case that the matrix $\mathbf{W}$ is shared among all layers, the decoupled case with different parameters in each layer is studied in~\cite{Chen2018theoretical} and linear convergence is further proved.

\subsection{Enhancing the PGD via Adaptive Depth}
Now consider the case of using the PGD method with a common parameter $R$ for recovering $K$ signals $\mathbf{x}_i$ from measurement vectors $\mathbf{y}_i=\mathbf{Ax}_i$ ($i=1,2,\ldots,K$). For the sake of exposition, we consider the noiseless case. If the value of $f(\mathbf{x}_i)$ is the same for all signals, the parameter could be tuned perfectly and set to $R=f(\mathbf{x}_i)$, which leads to the lowest upper bound for the reconstruction error according to Theorem~\ref{thm:stability_PGD}. However, if the value of $f(\mathbf{x}_i)$ varies for different signals, we cannot hope to recover all the signals exactly with a common parameter $R$, no matter how many iterations are conducted. This is a fundamental limitation of the classical PGD method for the case of the mismatch of $R$ and $f(\mathbf{x}_i)$. In this subsection, we demonstrate that one could overcome this drawback by applying adaptive depth (i.e., the number of iterations) for different signals.

\par
Without loss of generality, we assume $f(\mathbf{x}_1)< \ldots < f(\mathbf{x}_K)$ and an oracle PGD with adaptive depth, which conducts (\ref{eq:PGD_iteration}) with a parameter $R=f(\mathbf{x}_i)$ in the $t$th iteration satisfying $\sum_{j=0}^{i-1}\tau_{j}< t\leq \sum_{j=0}^{i}\tau_j$, and ejects the recovered signal of $\mathbf{x}_i$ immediately after $\sum_{j=0}^{i}\tau_j$ iterations, where $\tau_j$ ($j=1,\ldots,K$) are positive integers and $\tau_0=0$.

\begin{theorem}\label{thm:PGD_adaptive_general}
Let $f:\mathbb{R}^m\rightarrow \mathbb{R}$ be a norm function, $\mathcal{K} = \{\mathbf{x}\in \mathbb{R}^m:\ f(\mathbf{x})\leq R\}$ be a set, $\mathbf{A}\in \mathbb{R}^{n\times m}$ have independent $\mathcal{N}(\mathbf{0},1)$ entries, $\mathbf{x}_i\in \mathbb{R}^m$ ($i=1,\ldots,K$) be arbitrary vectors with $f(\mathbf{x}_1)< \ldots < f(\mathbf{x}_K)$, and $\mathbf{y}_i=\mathbf{Ax}_i\in\mathbb{R}^n$. Let $\kappa_f$ be a constant that is equal to 1 for convex function $f$ and equal to 2 for non-convex function $f$. Set the learning parameter to $\beta=\frac{1}{2}\left(\frac{\Gamma(\frac{n}{2})}{\Gamma(\frac{n+1}{2})}\right)^2\approx \frac{1}{n}$. Let $n_i$ be the minimum number of measurements required by the phase transition curve in (\ref{eq:Phase_Transition}), and $\rho_i=\sqrt{8\kappa_f^2\frac{n_i}{n}}$. Assume $n>8\kappa_f^2 n_i$ for all $i$, and all algorithms start from the initial point $\mathbf{x}_0=\mathbf{0}$. Then the total reconstruction error of the oracle PGD approach obeys
\begin{equation}\label{eq:PGD_general13}
\lim_{t\rightarrow \infty}\sum\limits_{i=1}^{K}\|\mathbf{x}_{t,i}-\mathbf{x}_i\|_2=0,
\end{equation}
with probability at least $1-8Le^{-\frac{\eta^2}{8}}$, where $\eta$ is a parameter controlling the probability of success and is defined in Definition 3 (Phase Transition Function), and $t$ is the iteration index.
\end{theorem}

\begin{IEEEproof}
We first derive the upper bound of the reconstruction error of the $i$th signal, i.e., $\mathbf{x}_i$. The oracle PGD ejects the recovered signal immediately after $\sum_{j=0}^{i}\tau_j$ iterations. For the $t$th iteration satisfying $\sum_{j=0}^{i-1}\tau_j< t\leq \sum_{j=0}^{i}\tau_j$, the oracle PGD conducts (\ref{eq:PGD_iteration}) with $R=f(\mathbf{x}_i)$. Define $\mathcal{K}_i = \{\mathbf{x}\in \mathbb{R}^m:\ f(\mathbf{x})\leq f(\mathbf{x}_i)\}$. According to Theorem \ref{thm:stability_PGD}, the reconstruction error of $\mathbf{x}_i$ obeys
\begin{equation}\label{eq:PGD_general_p1}
\begin{split}
\|\hat{\mathbf{x}}_{i}-\mathbf{x}_i\|_2\leq\ &\rho_i^{\sum_{j=0}^{i}\tau_j}\|\mathbf{x}_i\|_2+\\
&\sum_{h=1}^{i-1}\rho_i^{\sum_{j=0,j\neq h}^{i}\tau_j}
\frac{3-\kappa_f}{1-\rho_i}\|\mathbf{x}_i-\mathcal{P}_{\mathcal{K}_h}(\mathbf{x}_i)\|_2,
\end{split}
\end{equation}
with probability at least $1-8e^{-\frac{\eta^2}{8}}$.
If $\tau_i\rightarrow \infty$, the right-hand-side of (\ref{eq:PGD_general_p1}) tends to be zero. Therefore, for all the $K$ signals with $\tau_i\rightarrow \infty$ ($i=1,\ldots,L$), the total reconstruction error of the oracle PGD approach obeys
\begin{equation}\label{eq:PGD_general_p2}
\lim_{t\rightarrow \infty}\sum\limits_{i=1}^{K}\|\mathbf{x}_{t,i}-\mathbf{x}_i\|_2=0.
\end{equation}
\end{IEEEproof}

\emph{Remark:} Theorem \ref{thm:PGD_adaptive_general} states that a set of signals with different constraints, e.g., sparsity levels, can be successfully recovered by one PGD algorithm with fixed parameters and adaptive depth. Without adaptive depth, the upper bound of the total reconstruction error cannot approach zero\footnote{Here, we consider an oracle PGD with fixed parameters, which recovers different signals one by one.} even $t\rightarrow \infty$, as there will always exist mismatch between the parameter $R$ and different $f(\mathbf{x}_i)$.

\par
Existing DL based approaches for solving the sparse linear inverse problem either focus on the case of a fixed sparsity level, or ignore the diversity of sparsity levels of signals and hope the learned neural network is able to handle this divergence in a ``black box'' manner. Theorem 2 and 3 show that the PGD with learned gradient and adaptive depth is better than the traditional methods\footnote{Although the above oracle PGD algorithm requires the knowledge of the sparsity level for all tasks, which may not be practical, the above analysis unveils the potential benefits provided by our proposed method in the sequel.}, which provides the motivations for the neural network design with adaptive depth in later sections.


\begin{figure*}[!t]%
\centering%
\includegraphics[width=0.8\textwidth]{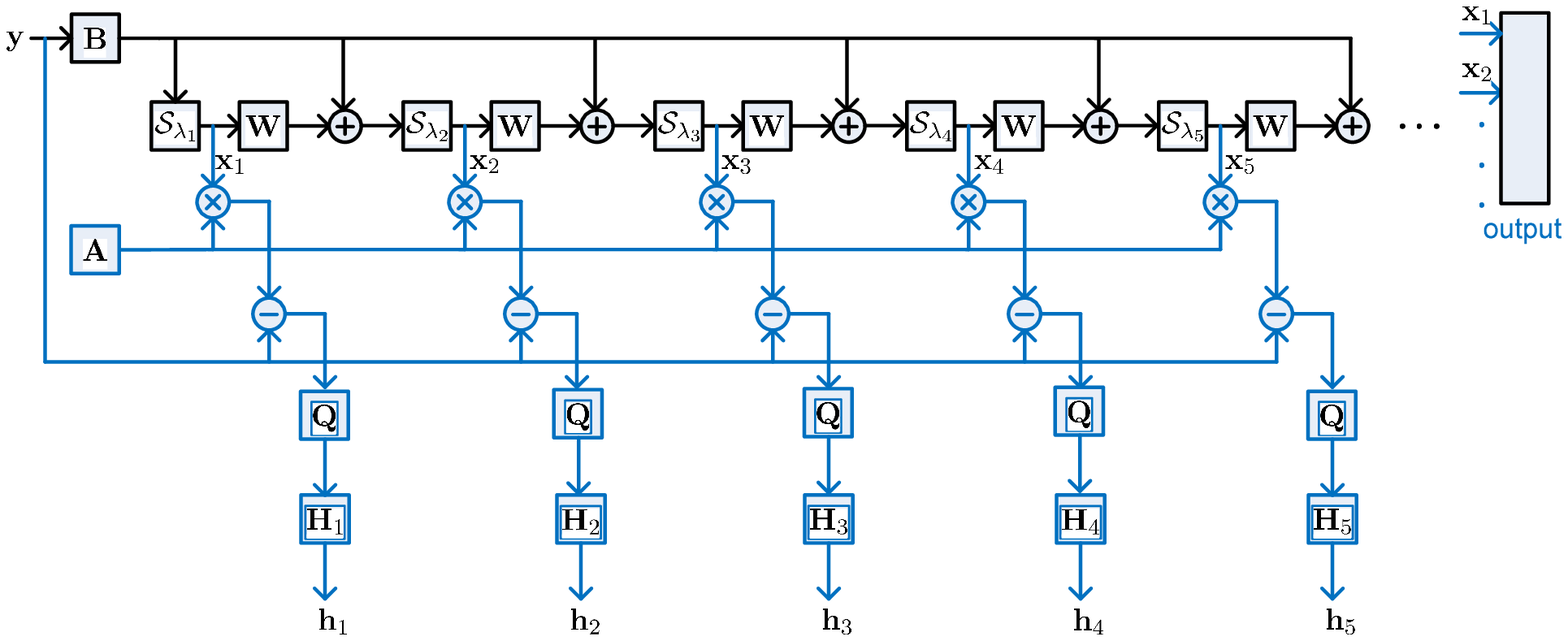}%
\DeclareGraphicsExtensions. \caption{The LISTA network with the proposed adaptive depth architecture. (Blue highlights the part of the structure that differs with the original LISTA.)} \label{fig:Adp_LISTA}
\end{figure*}%

\section{Enhancing the LISTA Network via Adaptive Depth}
In this section, we propose a novel method to enable neural networks to learn how many layers to execute to emit the output. We would like to clarify that the focus of this work is exploring the mechanism and architecture for adapting depth, which can be employed in many existing neural networks for solving sparse linear inverse problems~\cite{7934066,he2017bayesian,sun2016deep,andrychowicz2016learning,hershey2014deep,7442798,xin2016maximal,wang2016learning}. Here, we elaborate the proposed method in the context of the LISTA, which has a simple network structure and is closely related to the PGD.

\par
To begin with, we give a high level description of the proposed method. It has a nonsymmetric training-testing process. The proposed method modifies the conventional neural networks by adding a branch to the outputs of each layer, which predicts a halting score in the range $[0,1]$. In the training stage, the network is learned end-to-end, while in the inference stage we skip the remaining layers once the halting score reaches some given threshold. The proposed method is promising, as it requires small changes to the network structure and could be exploited in many different existing networks. The structure of LISTA with the proposed adaptive depth architecture is shown in Fig. \ref{fig:Adp_LISTA}.

\subsection{Halting Score}
In the proposed method, extra outputs, i.e., halting scores $h_t\in [0,1]$ ($1\leq t\leq L$), are added to each layer. Ideally, we would like the halting score to be a function of the reconstruction error $\|\mathbf{x}-\mathbf{x}_t\|_2^2$ and to indicate whether to eject the output in the inference. However, $\mathbf{x}$ is unknown and thus cannot be used as inputs or parameters of the network. To approximate the reconstruction error at each layer, one option is to consider $\|\mathbf{y}-\mathbf{Ax}_t\|_2^2$, which has been used as a stopping criterion in many existing iterative algorithms. The approximation accuracy highly depends on the interplay between the measurement matrix $\mathbf{A}$ and the error $\mathbf{x}-\mathbf{x}_t$. This brings up the question: is it possible to obtain a more accurate approximation of the reconstruction error $\|\mathbf{x}-\mathbf{x}_t\|_2^2$ and design a better indicator for making the halting decision.

\par
Here, we suggest to use an alternative design, i.e., $\|\mathbf{Q}(\mathbf{y}-\mathbf{Ax}_t)\|_2^2$, as the approximation of the reconstruction error in each iteration, where $\mathbf{Q}\in \mathbb{R}^{n\times n}$ denotes some linear mapping for the residual $\mathbf{y}-\mathbf{Ax}_t=\mathbf{A}(\mathbf{x}-\mathbf{x}_t)$. This design is inspired from the fact that the error $\mathbf{x}-\mathbf{x}_t$ is not an arbitrary vector in the underdetermined and structured case (i.e., sparse), otherwise the iterates of PGD in (\ref{eq:PGD_iteration}) could not converge~\cite{8107507}. Although it is difficult to characterize the distribution of the error $\mathbf{x}-\mathbf{x}_t$, we could learn the mapping matrix $\mathbf{Q}$ together with the algorithm (neural network). In this work, we design the halting score function as
\begin{equation}\label{eq:halting_score}
h_t=H_t(\mathbf{y},\mathbf{A},\mathbf{x}_t)=\sigma\left(\phi_t \|\mathbf{Q}(\mathbf{y}-\mathbf{Ax}_t)\|_2^2+\psi_t\right),
\end{equation}
where $\phi_t>0$ and $\psi_t$ are parameters of the layer $t$, $\mathbf{Q}$ is shared among all layers, and $\sigma(\cdot)$ denotes the sigmoid function that returns a value from 0 to 1. With an appropriate cost function, the mapping $\mathbf{Q}$ could be learned to take the error distribution into consideration, which leads to a more accurate approximation of $\|\mathbf{x}-\mathbf{x}_t\|_2^2$. Note that $\phi_t>0$ makes the halting score decreases with the reduction of the ``residual'' $\|\mathbf{Q}(\mathbf{y}-\mathbf{Ax}_t)\|_2^2$.

\subsection{Cost Function}
Define $\boldsymbol{\theta}$ as the set of variables that includes all the parameters of the neural network. We put forth a \textbf{continuous} and \textbf{differentiable} cost function
\begin{equation}\label{eq:cost_function_1}
\mathcal{L}(\boldsymbol{\theta})=\sum_{t=1}^{L}\frac{\|\mathbf{x}-\mathbf{x}_t\|_2^2}{h_t}+\tau h_t,
\end{equation}
where $\tau\geq0$ is a regularization parameter, both $\mathbf{x}_t$ and $h_t$ are functions of $\boldsymbol{\theta}$, and $\mathbf{x}$ is the output of the neural network and is known in the training. Note that the number of layers $L$ is fixed in the training, while an output can be emitted at some layer when the halting condition is satisfied in the inference, which will be introduced shortly. The reconstructed signals $\mathbf{x}_t$ and the halting scores $h_t$ of all the layers have contributions in the cost function. The cost function (\ref{eq:cost_function_1}) is different to the cost function in traditional iterative algorithms and related unfolded DL approaches, where intermediate updates have no explicit contribution to the cost function.

\par
Now we give the connection of the proposed method with the previous analysis of PGD. Here, we consider a simplified case by letting $\mathbf{W}=\mathbf{I}-\beta\mathbf{B}\mathbf{A}$, which degrade the LISTA to the learned PGD as in (\ref{eq:Learned_PGD_iteration}). This simplification reduces the capacity of the LISTA, while facilitates the analysis to shed light on the key insight of the proposed method. According to (\ref{eq:learned_PGD2}) in the Theorem \ref{thm:learned_PGD}, the degraded cost function is upper bounded by
\begin{equation}\label{eq:learned_PGD2_0}
\begin{split}
\mathcal{L}(\boldsymbol{\theta})\leq \tau h_t + \sum_{t=1}^{L}&\frac{(\kappa_f\rho(\mathbf{B}))^t}{h_t}\|\mathbf{x}\|_2+\\
&\kappa_f\xi(\mathbf{B})\frac{1-(\kappa_f\rho(\mathbf{B}))^t}{(1-\kappa_f\rho(\mathbf{B}))h_t}\|\boldsymbol{\omega}\|_2.
\end{split}
\end{equation}
By minimizing this cost function, parameter $\mathbf{B}$ with small $\rho(\mathbf{B})$ and $\xi(\mathbf{B})$ would be favored, which leads to a small reconstruction error $\|\mathbf{x}_t-\mathbf{x}\|_2$ for the output of each layer.

\par
Nextly, we unveil the adaptive depth mechanism brought by the new cost function. By letting the derivative of the cost function regarding to $h_t$ to be zero, the learned optimal halting score is
\begin{equation}\label{eq:derivative_function_pro}
h_t= \frac{\|\mathbf{x}-\mathbf{x}_t\|_2}{\sqrt{\tau}}.
\end{equation}
According to (\ref{eq:derivative_function_pro}), a well-trained network would generate halting scores proportional to the reconstruction error of each layer. According to Theorem \ref{thm:learned_PGD}, the reconstruction error $\|\mathbf{x}_t-\mathbf{x}\|_2$ of the intermediate output of the network tends to decrease with the increase of $t$, if both $\rho(\mathbf{B})$ in (\ref{eq:learned_PGD3}) and $\xi(\mathbf{B})$ in (\ref{eq:learned_PGD4}) are sufficiently small. Then the halting score $h_t$ will also decrease with the increase of $t$. In the inference phase, a large $h_t$  will make the algorithm continue to conduct the next iteration/layer. Therefore, this cost function design is key to fulfill the adaptive depth mechanism.

\par
Furthermore, we would like to emphasize that the reconstruction error $\|\mathbf{x}-\mathbf{x}_t\|_2^2$ communicates with the update of the halting score $h_t$, which makes learning the mapping $\mathbf{Q}$ in the halting score function affected by the error distribution of the output of each layer. The proposed network can be trained in an end-to-end manner, and parameters of all layers are updated via stochastic gradient decent and back-propagation. By fixing $h_t=1$ ($t=1,\ldots,L-1$) and $h_L=0$, the new network is no different to the standard LISTA, and thus can be seen as a generalization of the LISTA. Although the proposed neural network has fixed depth in the training phase, it produces a halting score at each layer, which enables adaptive depth in the inference phase.

\par
The parameter update steps for the proposed loss function (\ref{eq:cost_function_1}) are given in the Appendix.

\subsection{Halting Condition in Inference}
In the inference phase, an output is ejected with an adaptive number of layers. The number of executed layers $T$ is determined as the index of the first unit where the halting score is smaller than $\varepsilon$
\begin{equation}\label{eq:halting_score_prop}
T=\min\left\{t:h_{t}\leq \varepsilon\right\},
\end{equation}
where $\varepsilon$ is a small constant that allows computation to halt. If the condition in (\ref{eq:halting_score_prop}) does not hold for all layers, the number of executed layer is $L$.

\par
Changing the value of the parameter $\varepsilon$ leads to a varying number of executed layers and also a varying reconstruction error. The halting score relates to the reconstruction error and regularization parameter $\tau$. The derivative in (\ref{eq:derivative_function_pro}) becomes zero at $h_t=\sqrt{\frac{\|\mathbf{x}-\mathbf{x}_t\|_2^2}{\tau}}$. If one expect to terminate the computation at some layer with an error\footnote{Here, we assume the network has the capability to solve the sparse linear inverse problem.} $\|\mathbf{x}-\mathbf{x}_t\|_2^2=0.0001$ and set the regularization parameter $\tau=1$, then the derivative in (\ref{eq:derivative_function_pro}) becomes zero with $\varepsilon=0.01$, which gives us some guidance on selecting/tuning the parameter $\varepsilon$ in the halting condition. The computational complexity of the standard LISTA is $\mathcal{O}(Lnm)$ in inference, while the computational complexity of the new LISTA with adaptive depth is $\mathcal{O}(\bar{L}nm)$, where $\bar{L}\leq L$ is the number of executed layers.

\begin{figure}[!t]%
\centering%
\includegraphics[width=0.35\textwidth]{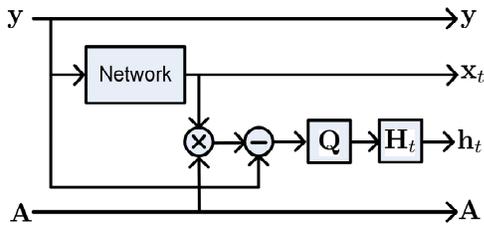}%
\DeclareGraphicsExtensions. \caption{The integration of the proposed adaptive depth architecture in the $t$th layer of an arbitrary network.} \label{fig:general}
\end{figure}%
\subsection{Extensions to Other Networks}
The proposed method can be easily incorporated into many existing neural networks for solving sparse linear inverse problems~\cite{BAI2020107729}. It requires few changes to the network structure and a small number of extra parameters. Fig.~\ref{fig:general} shows the integration of the proposed adaptive depth architecture in the $t$th layer of an arbitrary network. Note that $\mathbf{Q}\in \mathbb{R}^{n\times n}$ is shared for all layers, and $\phi_t$ and $\psi$ in the halting score function $H_t$ are unique in each layer.

\par
In addition to the adjustment in the network structure in Fig.~\ref{fig:general}, the cost function for existing neural networks also needs to be revised as (\ref{eq:cost_function_1}) to incorporate the impact of the halting score. In the inference, an output is ejected with an adaptive number of layers by evaluating the halting condition. These processes are the same as we described in the previous section for enhancing the LISTA.

\section{Experiments}
We evaluate the performance of the proposed architecture for adapting depth in solving the sparse linear inverse problem, where we consider numerical experiments with synthetic data and applications including random access in massive MTC and massive MIMO channel estimation.

\subsection{Learning Environment and Simulation Settings}
The training process is run on TensorFlow using a GTX1080Ti GPU. We use the Adam optimizer and the learning rate is set as $10^{-4}$ at the begining. We only reduce the learning rate if the smallest value of the loss does not change for $5,000$ mini-batches. To change the learning rate, we multiply it with a ratio smaller than 1. The used ratios are 0.1, 0.01 and 0.001, and the training is finished after all those ratios have already been used to update the learning rate. We consider the LISTA, the LISTA-CPSS~\cite{Chen2018theoretical} and the LAMP-L1~\cite{7934066} for comparison, although the proposed network structure can be added to various networks for enabling adaptive depth. To learn parameters in the proposed network with adaptive depth, we use a two-stage training process. Firstly, we initialize the part of the network parameters by using the learned parameters in its compared network which has fixed depth. In this stage, we lock these parameters and learn the parameters related to the halting score, which include $\{\phi_t,\psi_t\}$ for each layer $t$ and $\mathbf{Q}$ for all layers. In the second stage, we unlock and fine-tune all the parameters in the whole network. If not pointed out specifically in the experiments, the regularization parameter is set as $\tau=10$.

\subsection{Synthetic Experiments}
In this subsection, we conduct numerical investigation with synthetic data. A measurement matrix $\mathbf{A}$ of size $250\times 500$ is generated, where entries are generated independently from $\mathcal{N}(0,1)$ and then normalized for each column. For the generation of each $\mathbf{x}$, we first randomly select a sparsity level $s$ in the range between 10 and 100 with a uniform distribution, and then we generate a sparse vector $\mathbf{x}$ whose nonzero entries are drawn from $\mathcal{N}(0,1)$. Later we normalize $\mathbf{x}$, and obtain $\mathbf{y}$ using $\mathbf{y}=\mathbf{Ax}$. In the presence of measurement noise, we have $\mathbf{y}=\mathbf{Ax}+\mathbf{n}$, where elements in $\mathbf{n}$ are generated independently from $\mathcal{N}(0,1)$ and then scaled to satisfying some specified signal to noise ratio (SNR). In the training process, by repeating this process for $1000$ times, we obtain a mini-batch that includes $1000$ pairs of $\{\mathbf{x},\mathbf{y}\}$. We generate $300,000$ mini-batches for traning. In the inference process, the data samples are generated as described before, and each reported result is averaged over $10,000$ samples. The reconstruction error is defined as $\|\mathbf{x}-\mathbf{\hat{x}}\|_2^2$, where $\mathbf{\hat{x}}$ is the output of the network giving the input $\mathbf{x}$. Then the normalized mean square error (NMSE) is the average of $\frac{\|\mathbf{x}-\mathbf{\hat{x}}\|_2^2}{\|\mathbf{x}\|_2^2}$.


\begin{figure}[!t]
\centering
\subfigure[NMSE]{\includegraphics[width=0.45\linewidth]{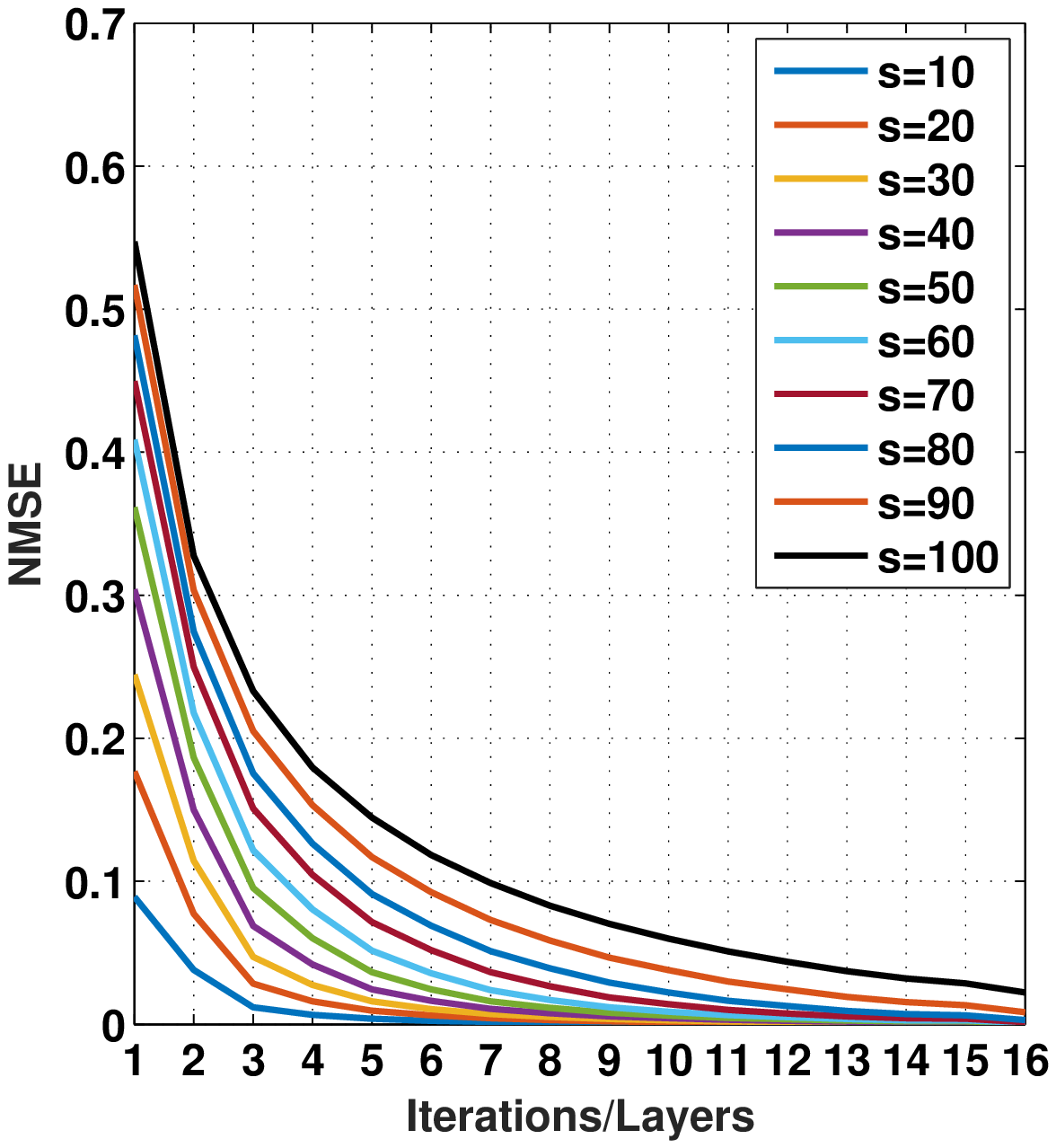}
\label{measurement}}
\subfigure[Halting scores]{\includegraphics[width=0.45\linewidth]{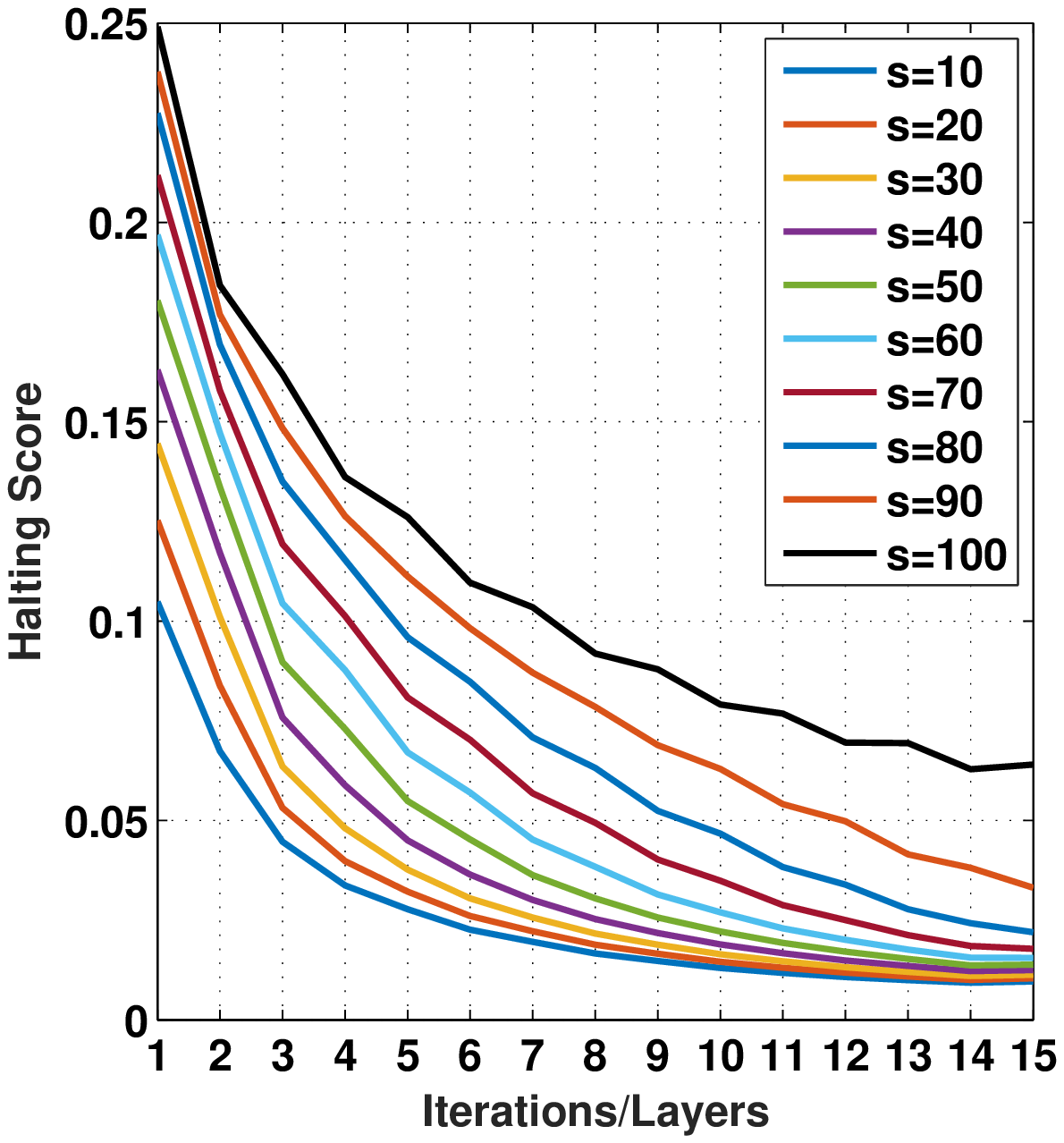}
\label{dimension}}
\caption{Comparison of the NMSE and the halting scores at different numbers of executing layers for the proposed LISTA with adaptive depth.} \label{fig:scores_layer}
\end{figure}

\begin{figure}[!t]
\centering
\subfigure[LISTA]{\includegraphics[width=0.45\linewidth]{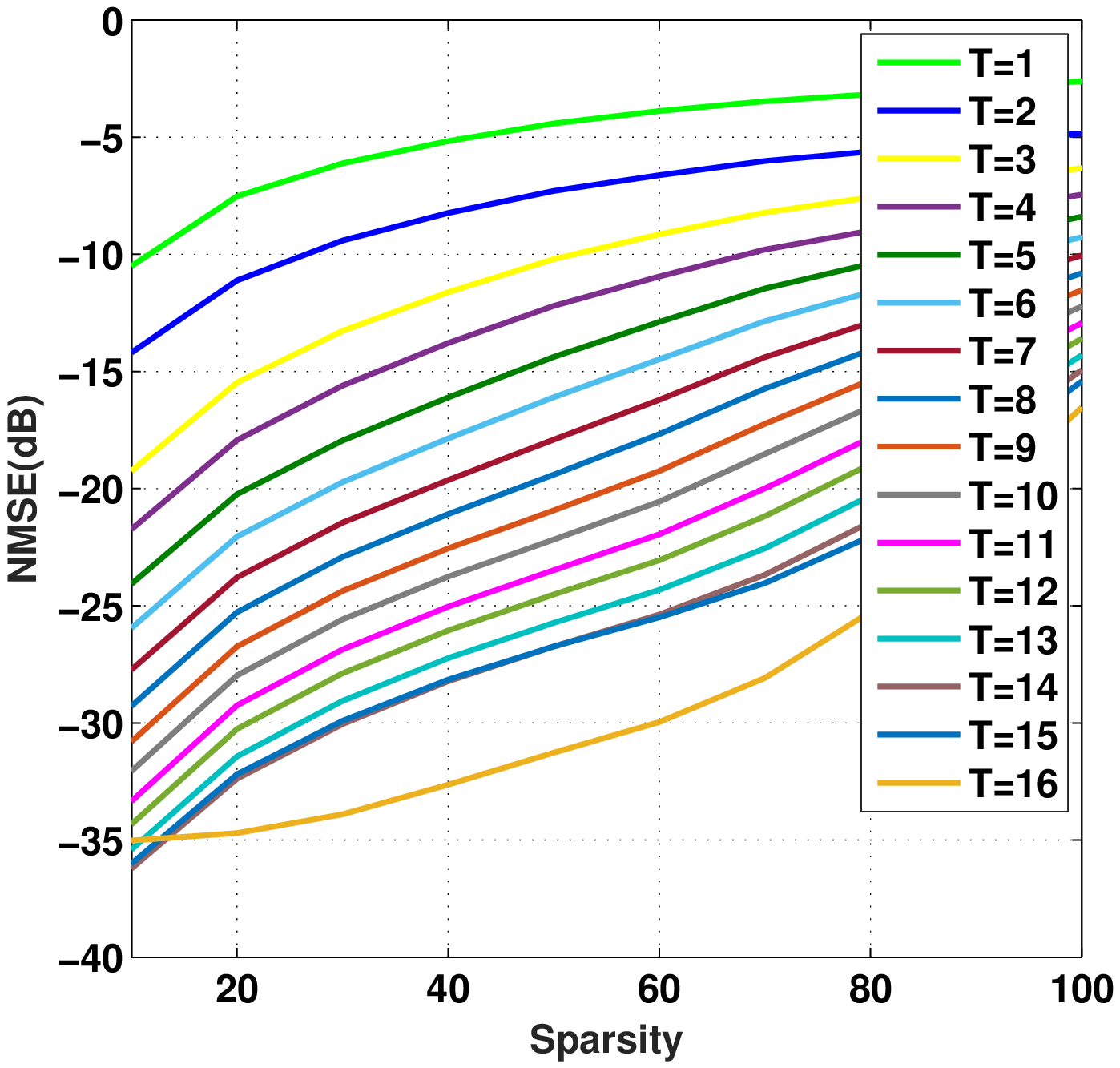}
\label{dimension}}
\subfigure[Proposed (NMSE)]{\includegraphics[width=0.45\linewidth]{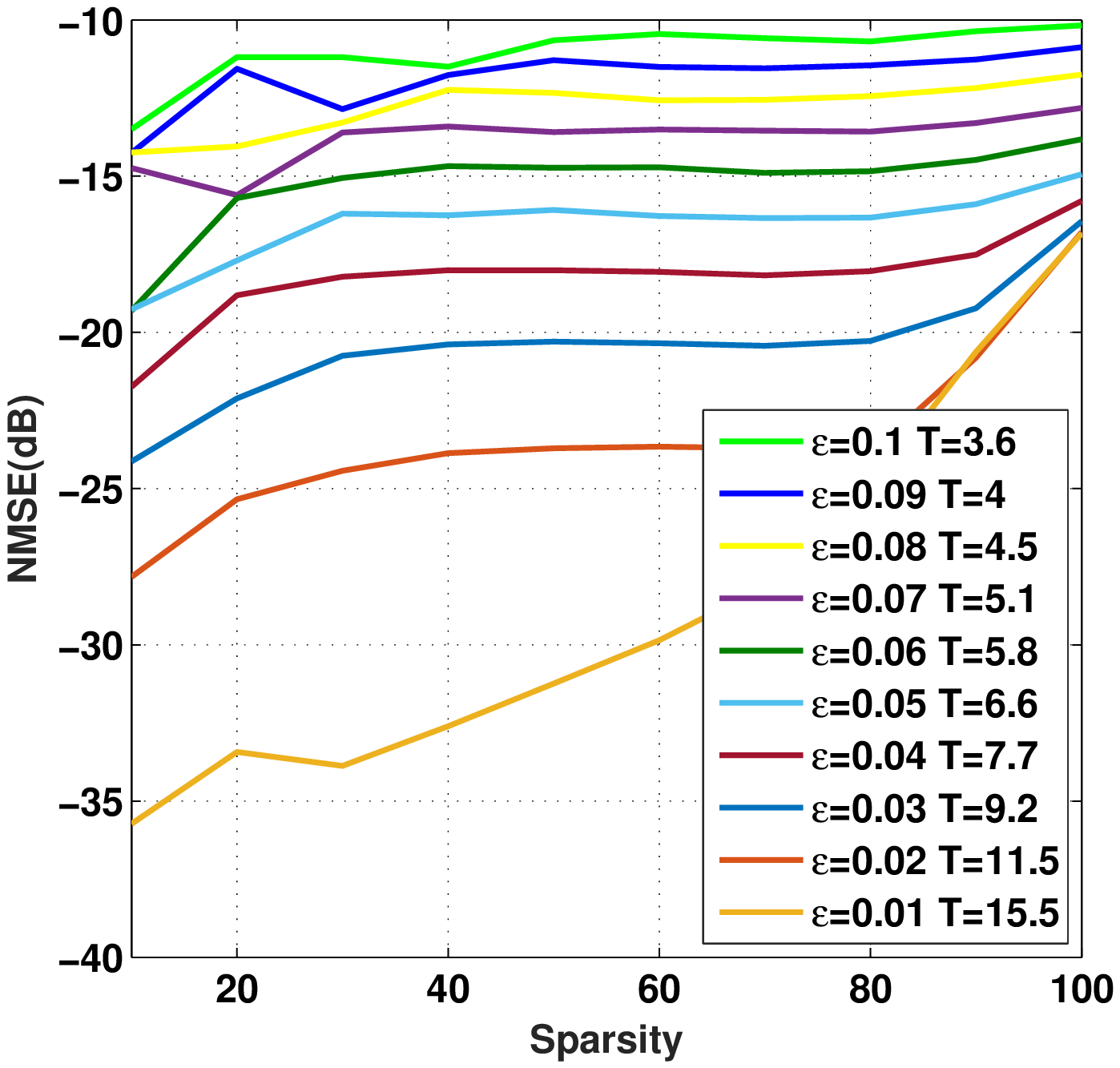}
\label{measurement}}
\subfigure[Proposed (executing layers)]{\includegraphics[width=0.45\linewidth]{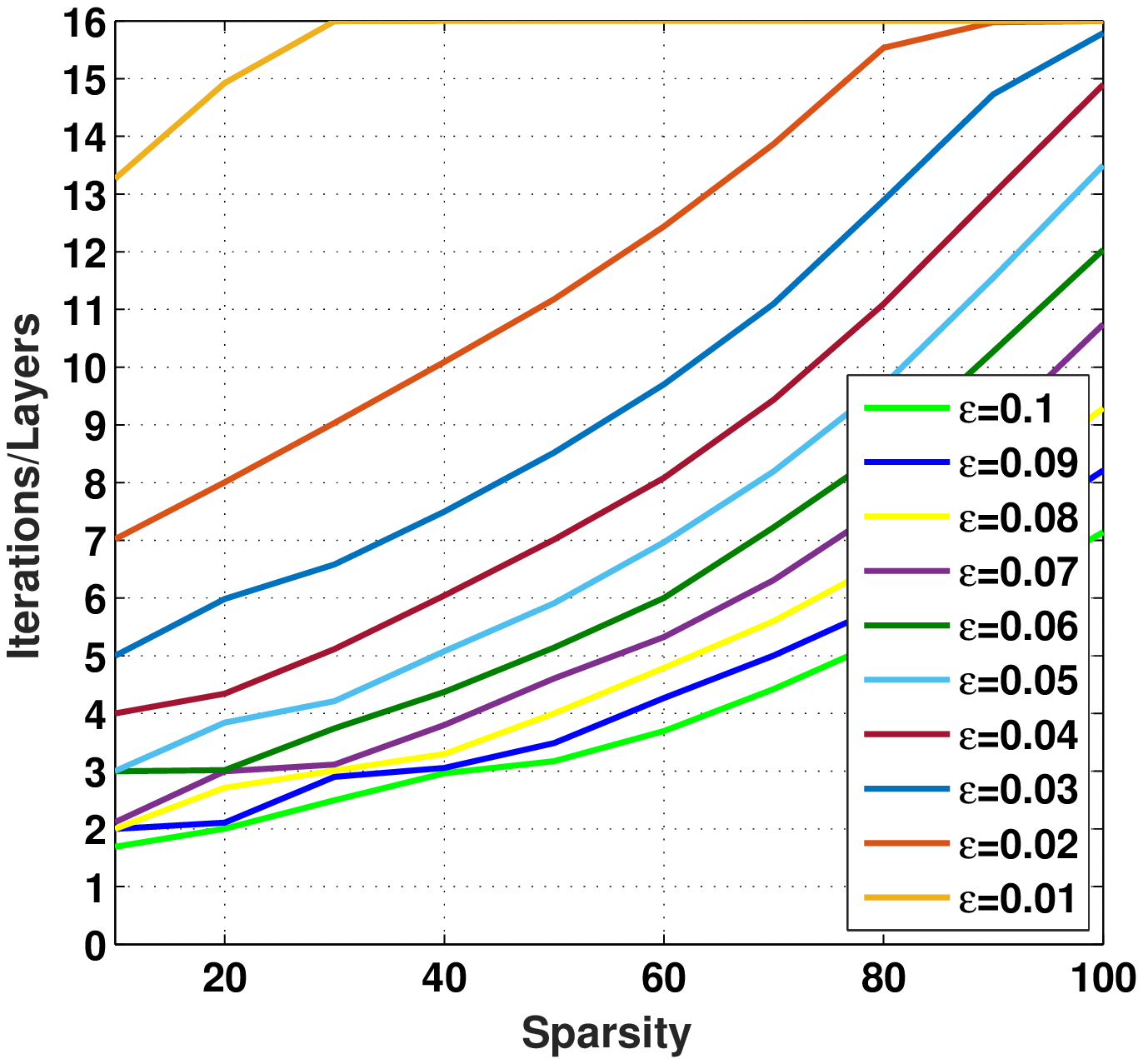}
\label{dimension}}
\caption{Comparison of the NMSE and the number of executing layers with different sparsity levels.} \label{fig:layer_sparsity}
\end{figure}

\par
In the first experiment, we evaluate the NMSE and the halting score at different executing layers of the proposed LISTA with adaptive depth. The result in Fig.~\ref{fig:scores_layer} shows that i) the NMSE and the halting score decreases when more layers are used; and ii) at the same layer, less sparse signals have lower NMSE and halting scores. By using some fixed halting constant $\varepsilon$ in (\ref{eq:halting_score_prop}), it will execute distinct numbers of layers to reconstruct signals of different sparsity levels.

\par
In the second experiment, we investigate further on how the NMSE and the number of executing layers adapt to the signal sparsity level in the proposed LISTA with adaptive depth. By varying $\varepsilon$ in (\ref{eq:halting_score_prop}), the network ejects the output at different layer $T$. As shown in Fig.~\ref{fig:layer_sparsity} (a) and (b), the proposed method with adaptive depth significantly reduces the performance difference of different sparsity levels. For example, the difference of the NMSE of the LISTA with 4 layers for sparsity level 10 and sparsity level 100 is $14.2dB$ according to Fig.~\ref{fig:layer_sparsity} (a), while the difference of the NMSE of the proposed LISTA with averaged 4 layers for sparsity level 10 and sparsity level 100 is only $3.4dB$ according to Fig.~\ref{fig:layer_sparsity} (b). In addition, it is observed that in Fig.~\ref{fig:layer_sparsity} (c), more layers of the network are activated when we use a small halting constant $\varepsilon$. Traditional DL based methods usually consider networks with fixed depth for all sparsity level, which differs with optimization based algorithms. The results in both Fig.~\ref{fig:scores_layer} and Fig.~\ref{fig:layer_sparsity} demonstrate the similarity of the proposed DL based method and optimization based algorithms in solving sparse linear inverse problems, where the computational complexity adapts to the hardness of the problem.

\begin{figure}[!t]
\centering
\subfigure[Noiseless case]{\includegraphics[width=0.45\linewidth]{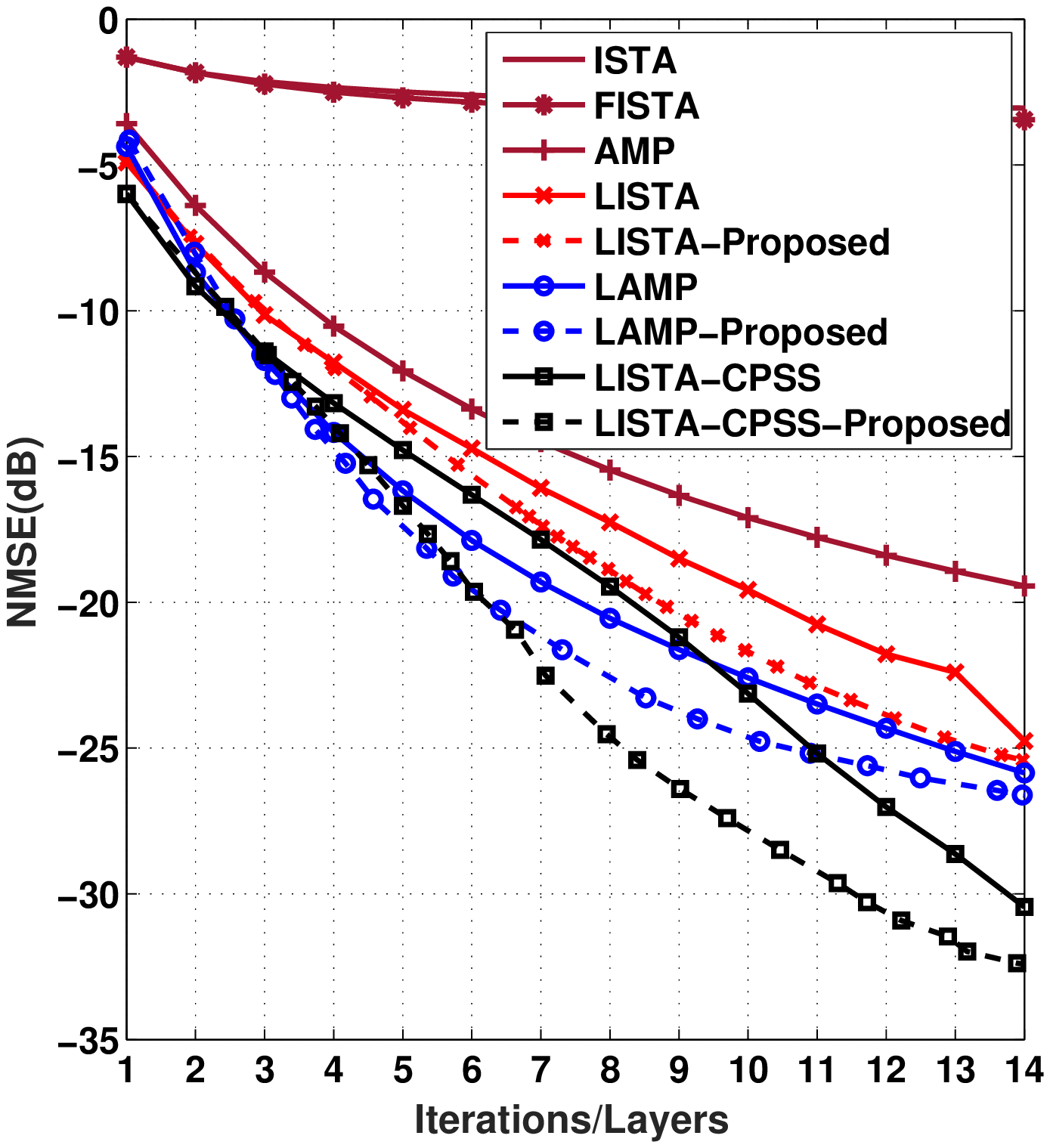}
\label{measurement}}
\subfigure[Noisy case (SNR=20dB)]{\includegraphics[width=0.45\linewidth]{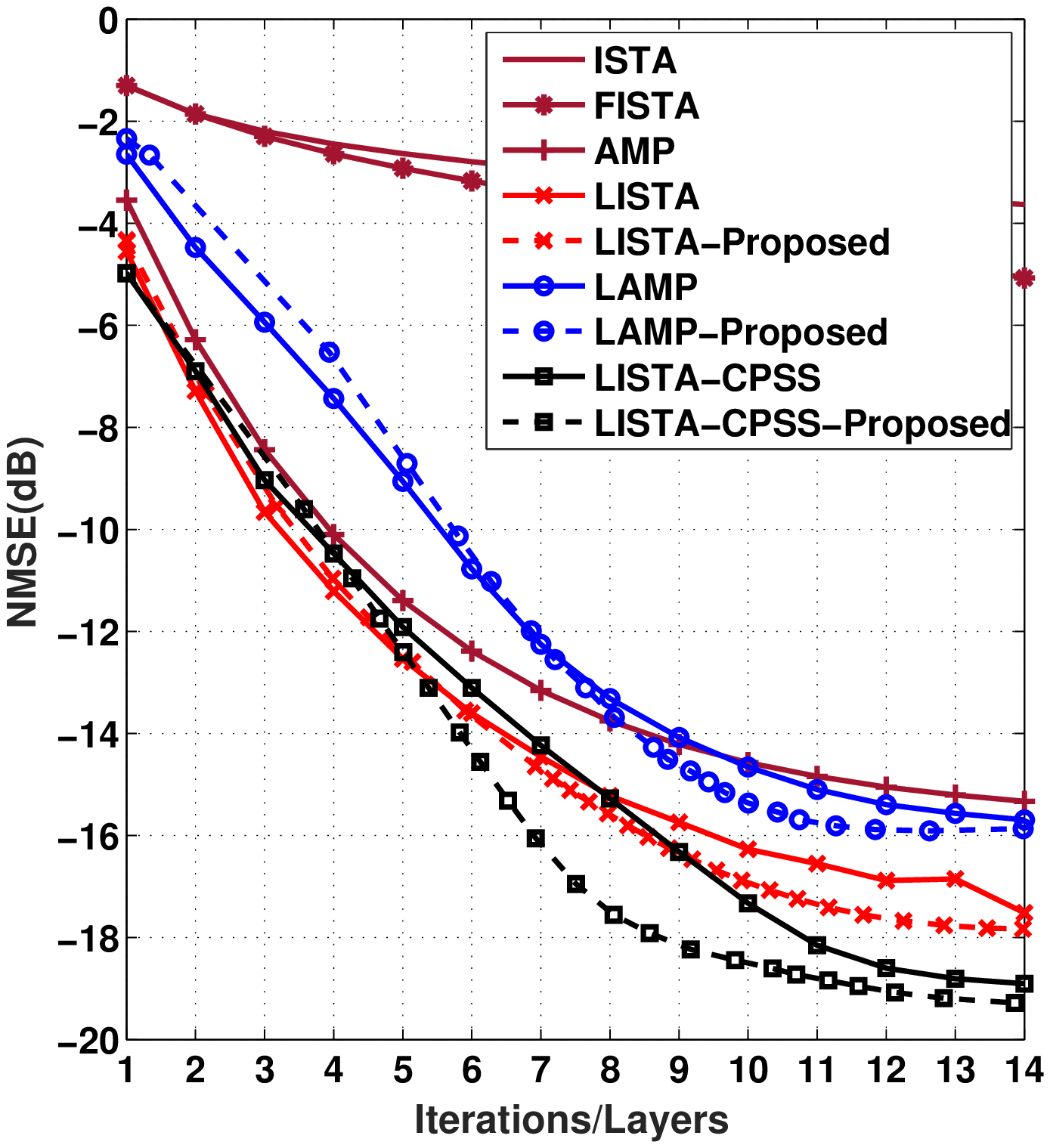}
\label{dimension}}
\caption{Comparison of the NMSE with different averaged numbers of executing layers.} \label{fig:error_layer}
\end{figure}
\begin{figure}[!t]
\centering
\subfigure[Noiseless case]{\includegraphics[width=0.45\linewidth]{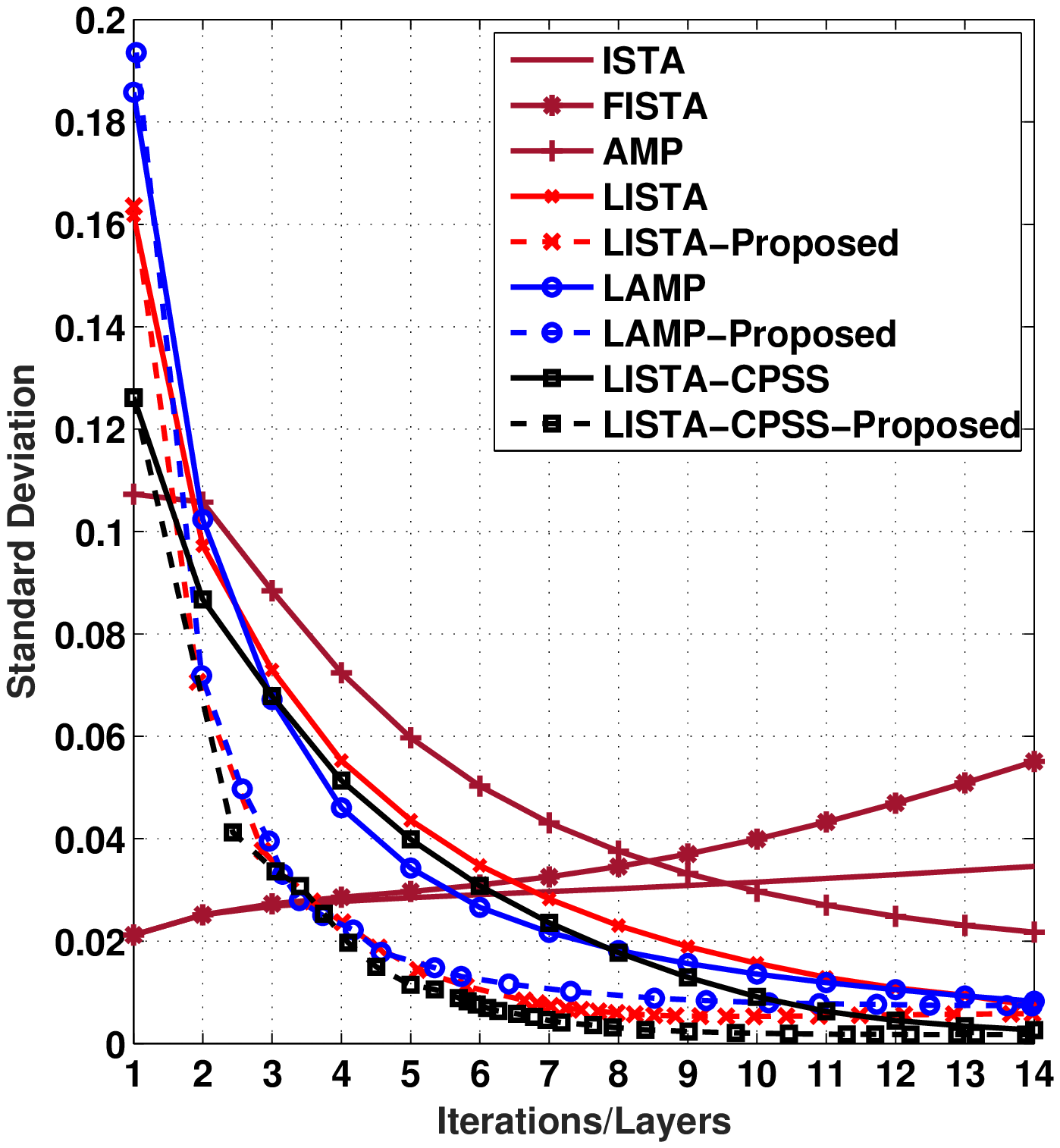}
\label{measurement}}
\subfigure[Noisy case (SNR=20dB)]{\includegraphics[width=0.45\linewidth]{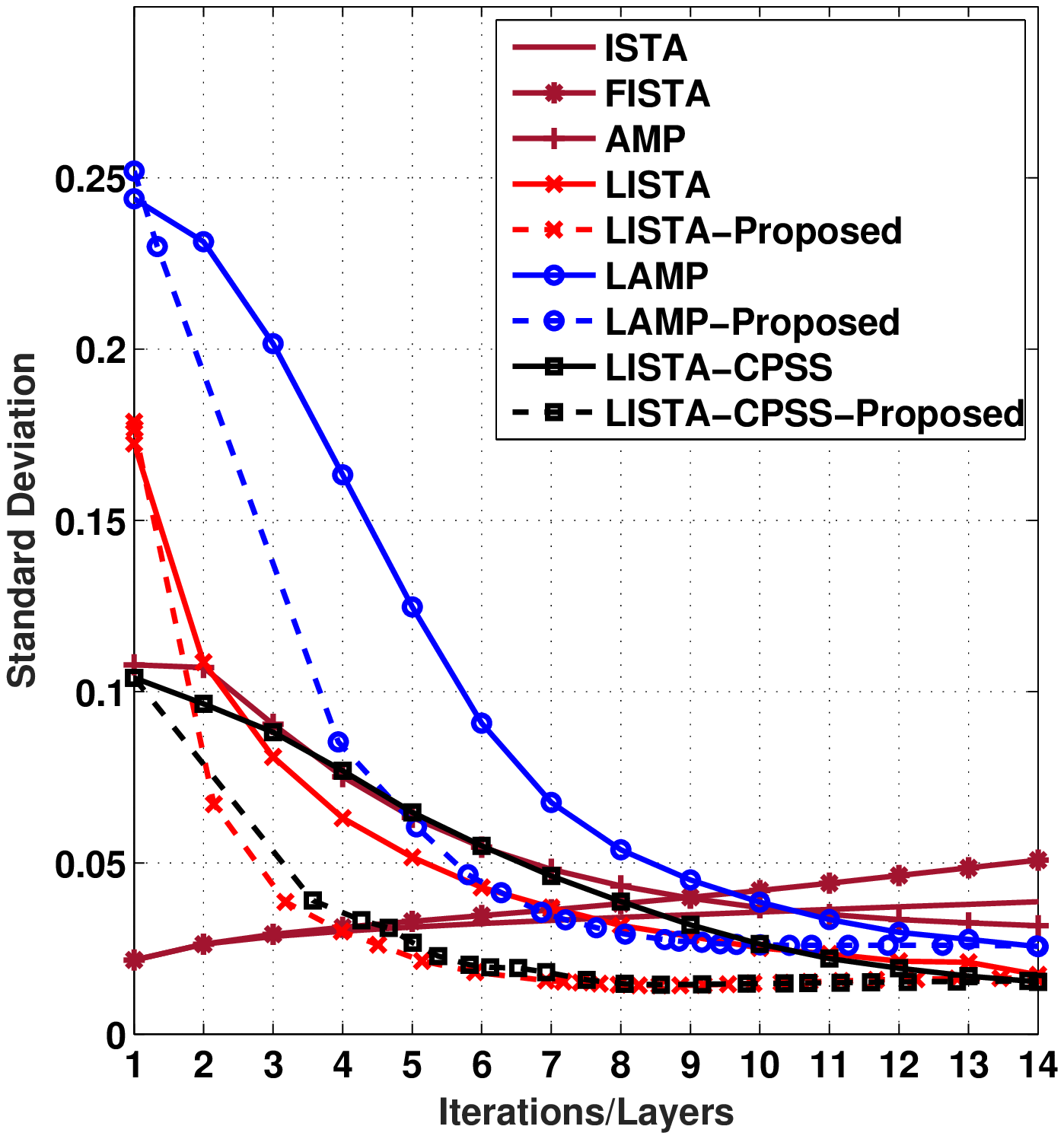}
\label{dimension}}
\caption{Comparison of the standard deviation of the reconstruction error with different averaged numbers of executing layers.} \label{fig:stdVSlayer}
\end{figure}
\begin{figure}[!t]
\centering
\subfigure[Noiseless case]{\includegraphics[width=0.45\linewidth]{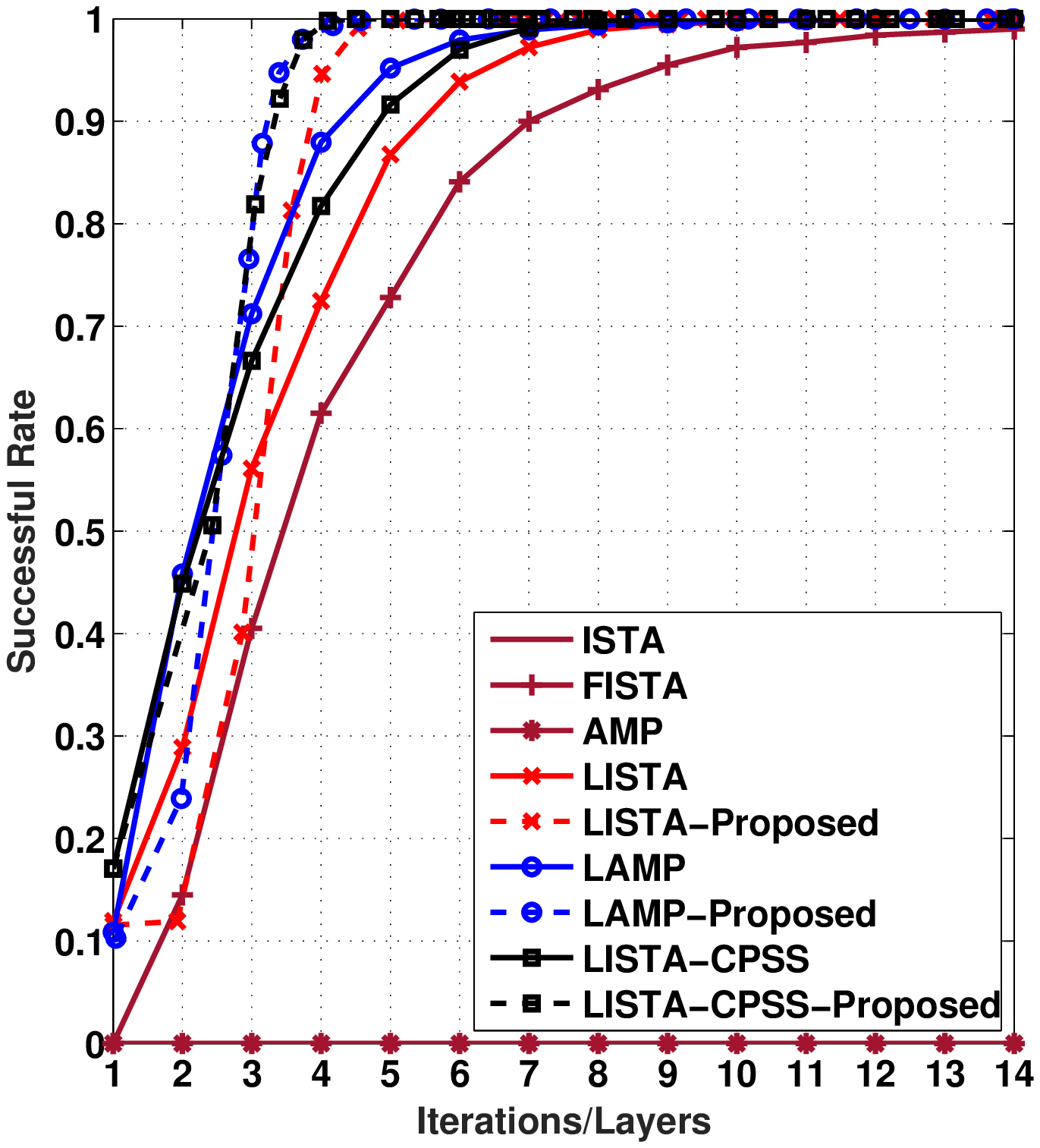}
\label{measurement}}
\subfigure[Noisy case (SNR=20dB)]{\includegraphics[width=0.45\linewidth]{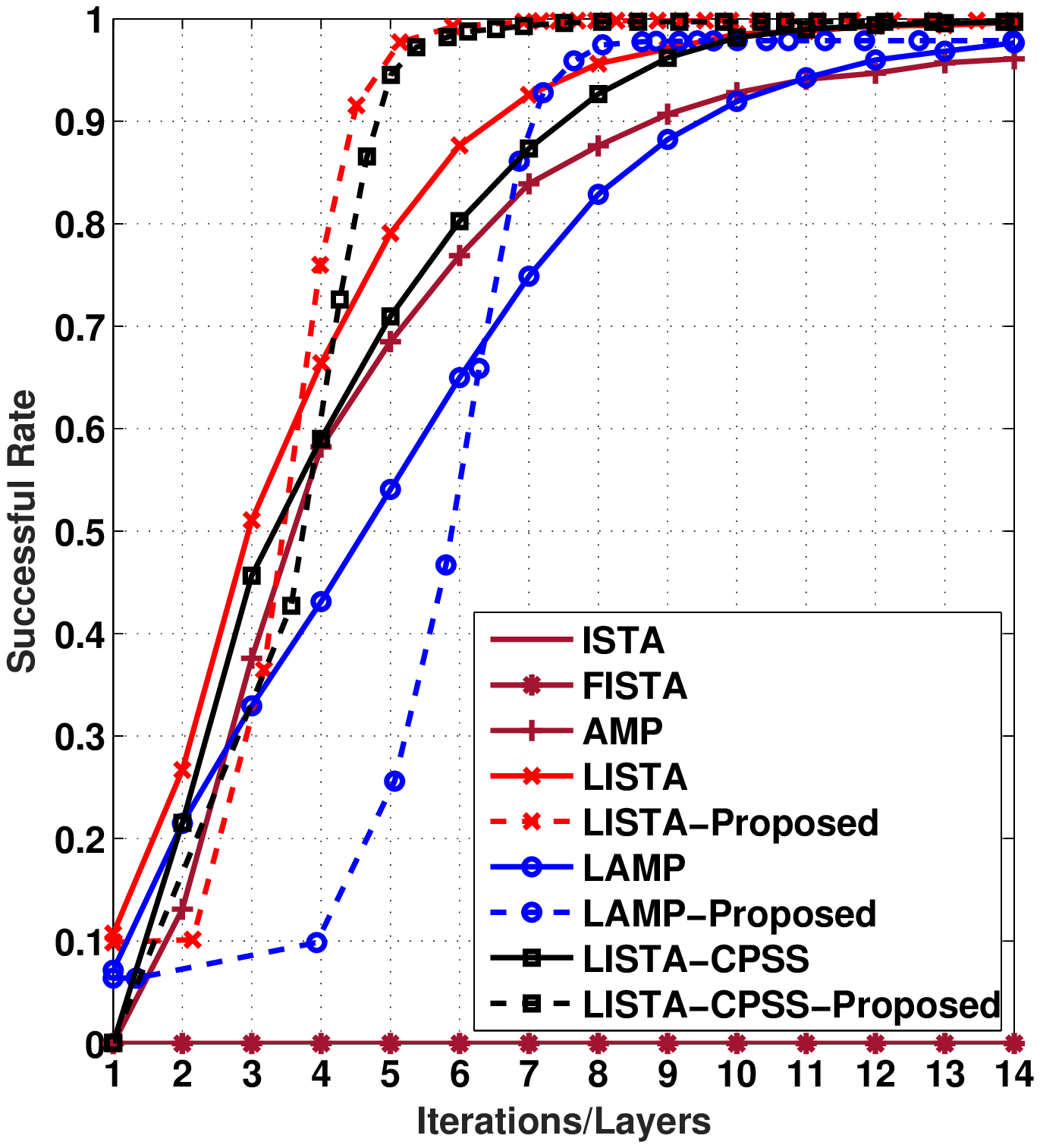}
\label{dimension}}
\caption{Comparison of the successful rate of the reconstruction with different averaged numbers of executing layers.} \label{fig:sucRate}
\end{figure}

\par
In the next experiment, we study the reconstruction performance (NMSE and error standard deviation) of the proposed method with different averaged numbers of executing layers. We train LISTA/LAMP/LISTA-CPSS networks of 14-layers, while set the the maximum depth of the proposed networks to be 16. By varying the value of the halting constant $\varepsilon$ in (\ref{eq:halting_score_prop}), the proposed networks eject outputs at different layers, which leads to varying reconstruction error. Then hard tasks can enjoy the benefit brought by extra layers in the proposed method, although the average number of layers used is no more than 14 (so that the comparison is fair). This setting is used in all the remaining experiments when traditional methods and proposed methods are compared. As shown in Fig.~\ref{fig:error_layer}, the proposed method significantly reduces the reconstruction error for the LISTA, the LAMP and the LISTA-CPSS, and in both the noiseless case and the noisy case (with a signal to noise ratio of 20dB). This result demonstrates the benefit brought by the proposed adaptive length structure, where easy tasks consumes less computation. We would like to emphasize that Fig.~\ref{fig:layer_sparsity} and Fig.~\ref{fig:error_layer} illustrate the depth-accuracy-hardness relationship of the proposed method, where the number of executing layers represents the depth, the reconstruction error (i.e., $\|\mathbf{x}-\mathbf{\hat{x}}\|_2^2$) represents the accuracy, and the sparsity level $s$ represents the hardness of the task. Furthermore, as shown in Fig.~\ref{fig:stdVSlayer}, the standard deviation of reconstruction error can be reduced by employing the proposed methods with adaptive depth. Not only the averaged performance but also the error standard deviation are important in determining the successful reconstruction rate. Assume the signal is successfully reconstructed if the NMSE is smaller than $-10$dB. Fig.~\ref{fig:sucRate} shows significant improvement of the successful reconstruction rate brought by the proposed method. For example, for the LISTA with and without adaptive depth, 5 layers and 9 layers are required to guarantee success reconstruction in the noiseless case, respectively, and 7 layers and 12 layers are required to guarantee success reconstruction in the noisy case, respectively. Note that when we let the halting score $\varepsilon$ be close to zero, all layers of the proposed network are active for all the sparsity levels. In this case, the proposed network is equivalent to the traditional network.

\begin{figure}[!t]
\centering
\subfigure[NMSE]{\includegraphics[width=0.45\linewidth]{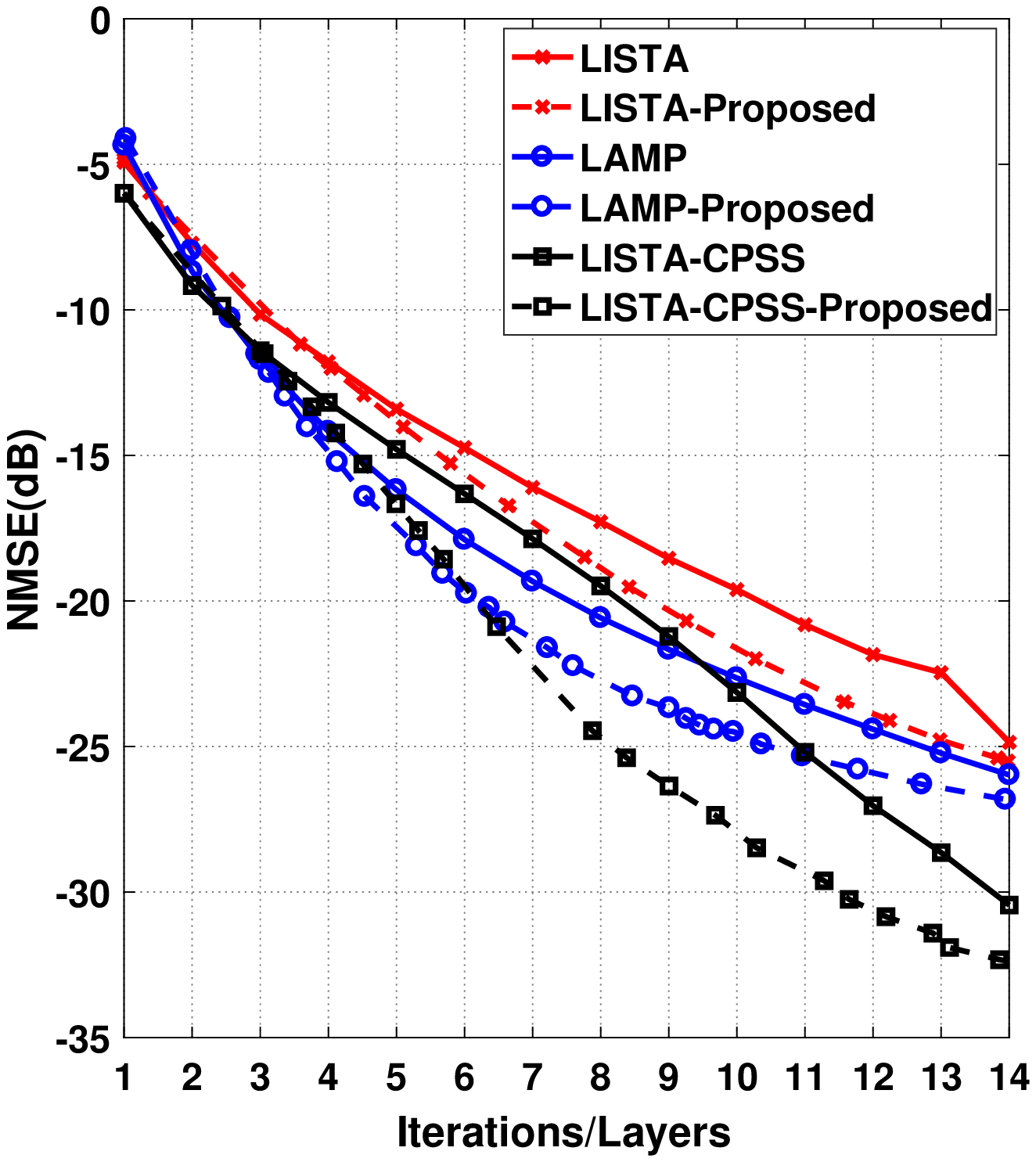}
\label{measurement}}
\subfigure[Successful Rate]{\includegraphics[width=0.45\linewidth]{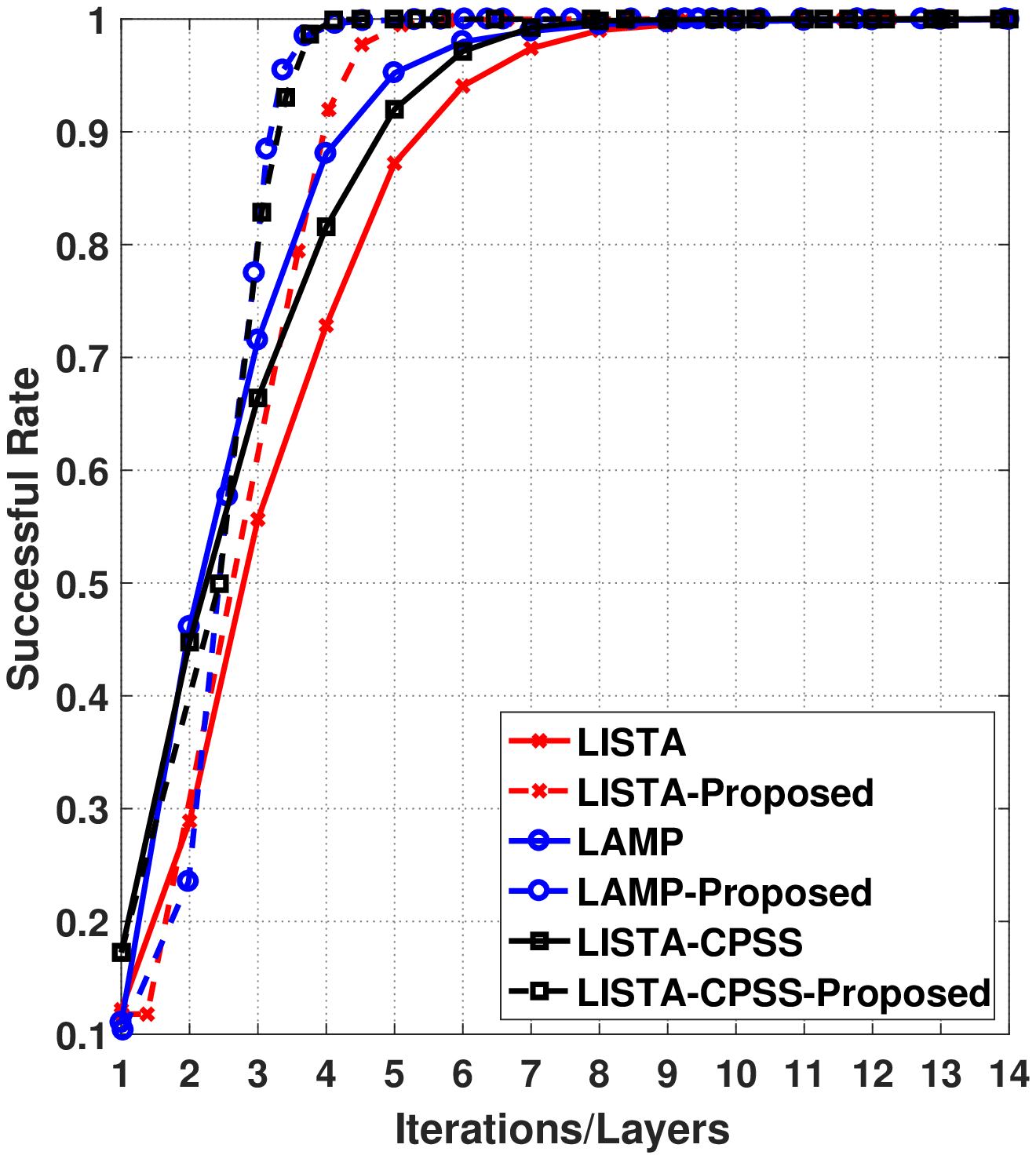}
\label{dimension}}
\caption{Performance comparison with the Rademacher measurement matrix.} \label{fig:ra}
\end{figure}
\par
In addition to the random Gaussian measurement matrix $\mathbf{A}$, we investigate the case of Rademacher measurement matrices without the additive noise. As shown in Fig.~\ref{fig:ra}, the proposed approach with adaptive depth also has superior performance, which suggests our approach is not restricted to the case of the random Gaussian measurement matrix, and can be applied to other measurement matrix designs.

\begin{figure}[!tb]%
\centering%
\includegraphics[width=0.45\textwidth]{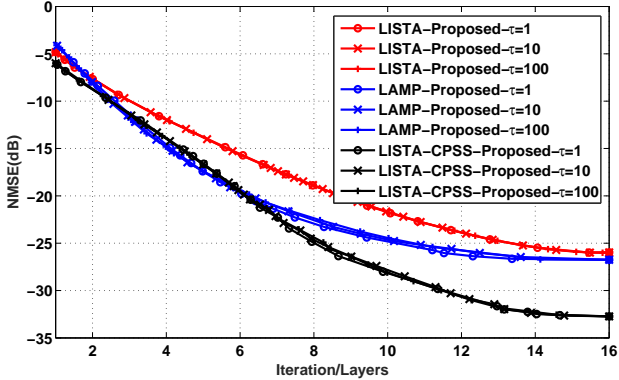}%
\DeclareGraphicsExtensions. \caption{Performance comparison with the regularization parameter of different values. } \label{fig:error_layer_tao}
\end{figure}%
\par
Now in Fig.~\ref{fig:error_layer_tao}, we provide experimental results to show how the regularization parameter $\tau$ in the cost function (\ref{eq:cost_function_1}) of the proposed method affects the trade-off performance between the reconstruction accuracy and the number of executing layers. It is observed that varying the regularization parameter $\tau=1,10,100$ has limited impact on the trade-off performance for all the three networks. Therefore, the result shows the proposed method is not very sensitive to the selection of the regularization parameter $\tau$.

\begin{figure}[!tb]%
\centering%
\includegraphics[width=0.45\textwidth]{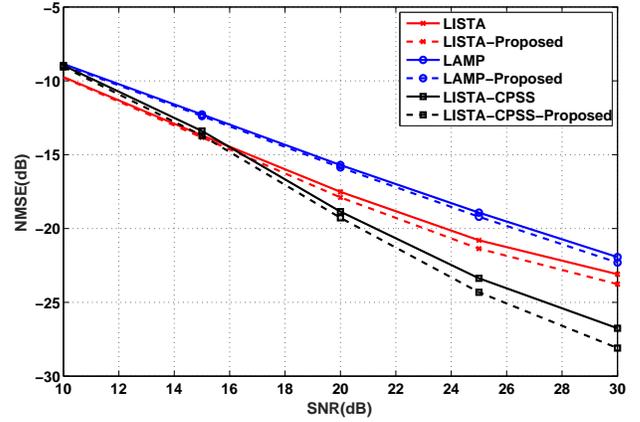}%
\DeclareGraphicsExtensions. \caption{Performance comparison with varying SNRs. } \label{fig:snr}
\end{figure}%

\par
Fig.~\ref{fig:snr} provides performance comparison for different SNRs. As in previous experiments, we train LISTA/LAMP/LISTA-CPSS networks of 14-layers,
while set the the maximum depth of the proposed networks to be 16. The results shown in Fig.~\ref{fig:snr} are the NMSE at the last layer, i.e., layer 14, of the network. It is observed that the gain of the proposed methods tends to increase with the grow of the SNR.

\par
It would be interesting to investigate different designs of the halting score prediction network. We consider three different designs including the design with a learned mapping matrix $\mathbf{Q}$ in (\ref{eq:halting_score}), the design without the mapping matrix $\mathbf{Q}$, i.e.,
\begin{equation}
h_{t}=\sigma(\phi_{t}\|(\mathbf{y}-\mathbf{Ax}_{t})\|_{2}^{2}+\psi_{t}),
\end{equation}
and the design using a two-layer fully connection neural network with $2n$ hidden neurons, i.e.,
\begin{equation}
h_{t}=\sigma(\mathbf{w}_{2t}^T\text{Relu}(\mathbf{W}_{1t}(\mathbf{y}-\mathbf{Ax}_{t})+\mathbf{b}_{1t})+{b}_{2t}),
\end{equation}
where $\mathbf{W}_{1t}\in \mathbb{R}^{2n\times n}$, $\mathbf{b}_{1t}\in \mathbb{R}^{2n}$, $\mathbf{w}_{2t}\in \mathbb{R}^{2n}$ and ${b}_{2t}\in \mathbb{R}$.
\begin{figure}[t]
\begin{center}
\includegraphics[width=0.9\linewidth]{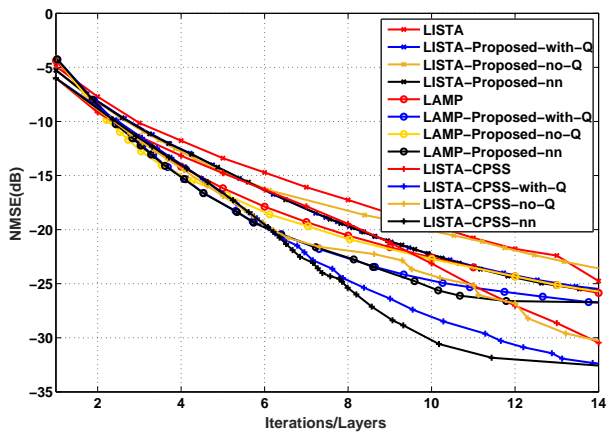}
\caption{Comparison of different halting network designs.}
\label{fig:prediction-network}
\end{center}
\end{figure}
\par
As illustrated in Fig~\ref{fig:prediction-network}, in comparison with the design without the mapping matrix $\mathbf{Q}$, our design leads to a significant performance improvement. This comparison showed the need of having an extra mapping $\mathbf{Q}$ to predict the halting score. Applying a more complex network to predict the halting score may not significantly improve the performance. For the LISTA, the proposed design with a simple mapping $\mathbf{Q}$ achieves the same performance as the design using two-layer neural network. For the LAMP and the LISTA-CPSS, the two-layer neural network shows some gain. However, the extra gain is small, and more parameters increase the computational complexity. Our design uses a shared $\mathbf{Q}\in \mathbb{R}^{n\times n}$ and layer-wise parameters $\phi_{t}$ and $\psi_{t}$. The total number of parameters of our design is only $n\times n+2L$, while the two-layer neural network has $(2n\times (n+2)+1)\times L$ parameters.


\begin{figure}[!tb]%
\centering%
\includegraphics[width=0.45\textwidth]{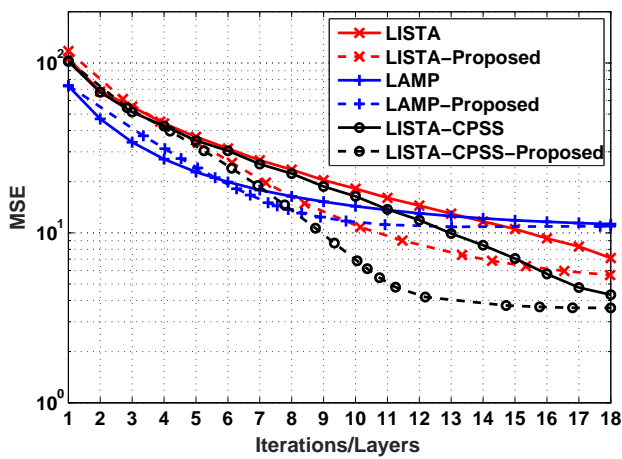}%
\DeclareGraphicsExtensions. \caption{Comparison of the reconstruction error with different averaged numbers of executing layers for CS-MUD in MTC. } \label{fig:random-access}
\end{figure}%

\begin{figure*}[t]
\centering
\subfigure[Distribution of sparsity]{\includegraphics[width=0.234\linewidth]{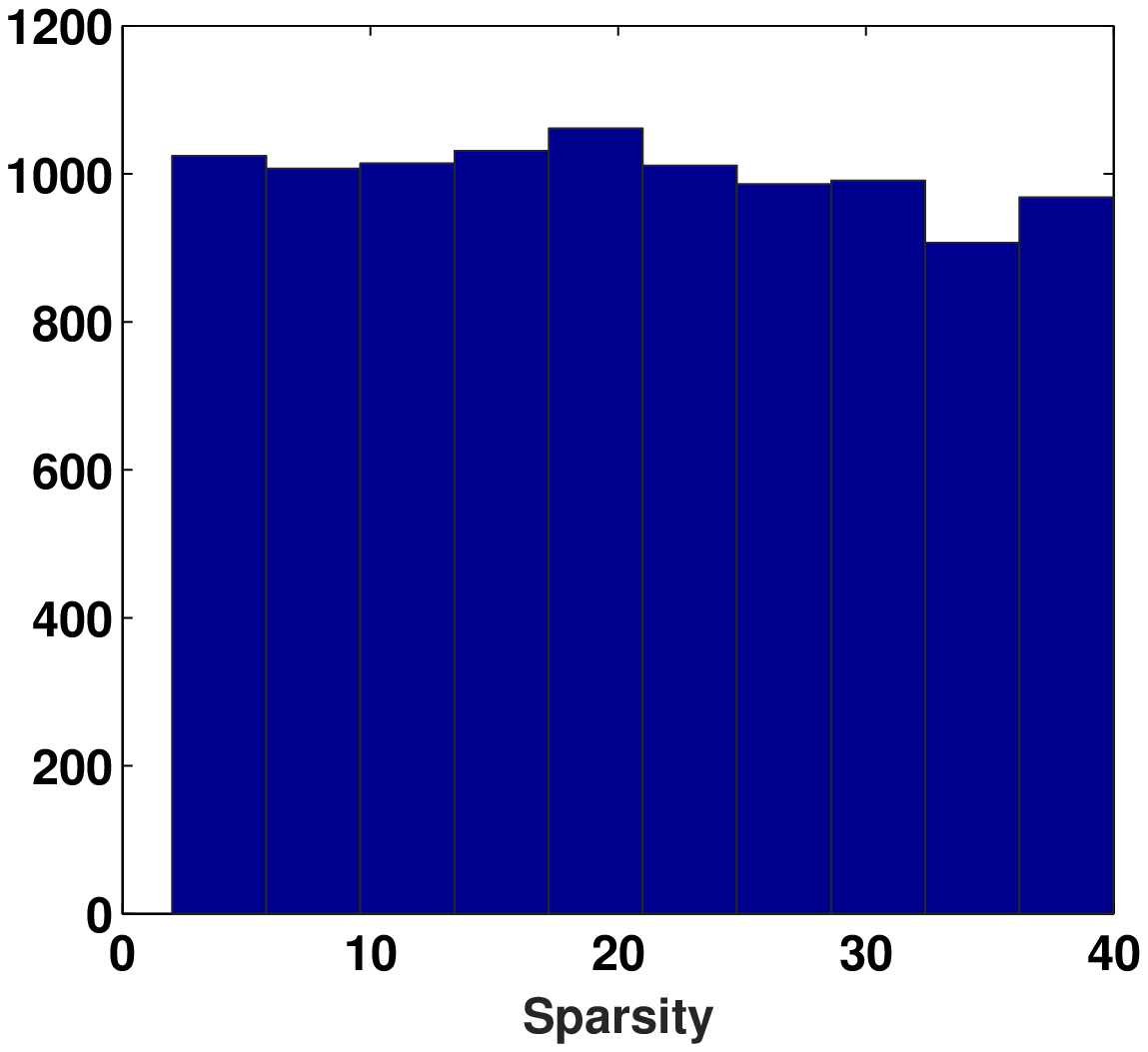}
\label{Dist-Sparsity}}
\subfigure[Distribution of the number of used layers ($\epsilon=0.1$)]{\includegraphics[width=0.234\linewidth]{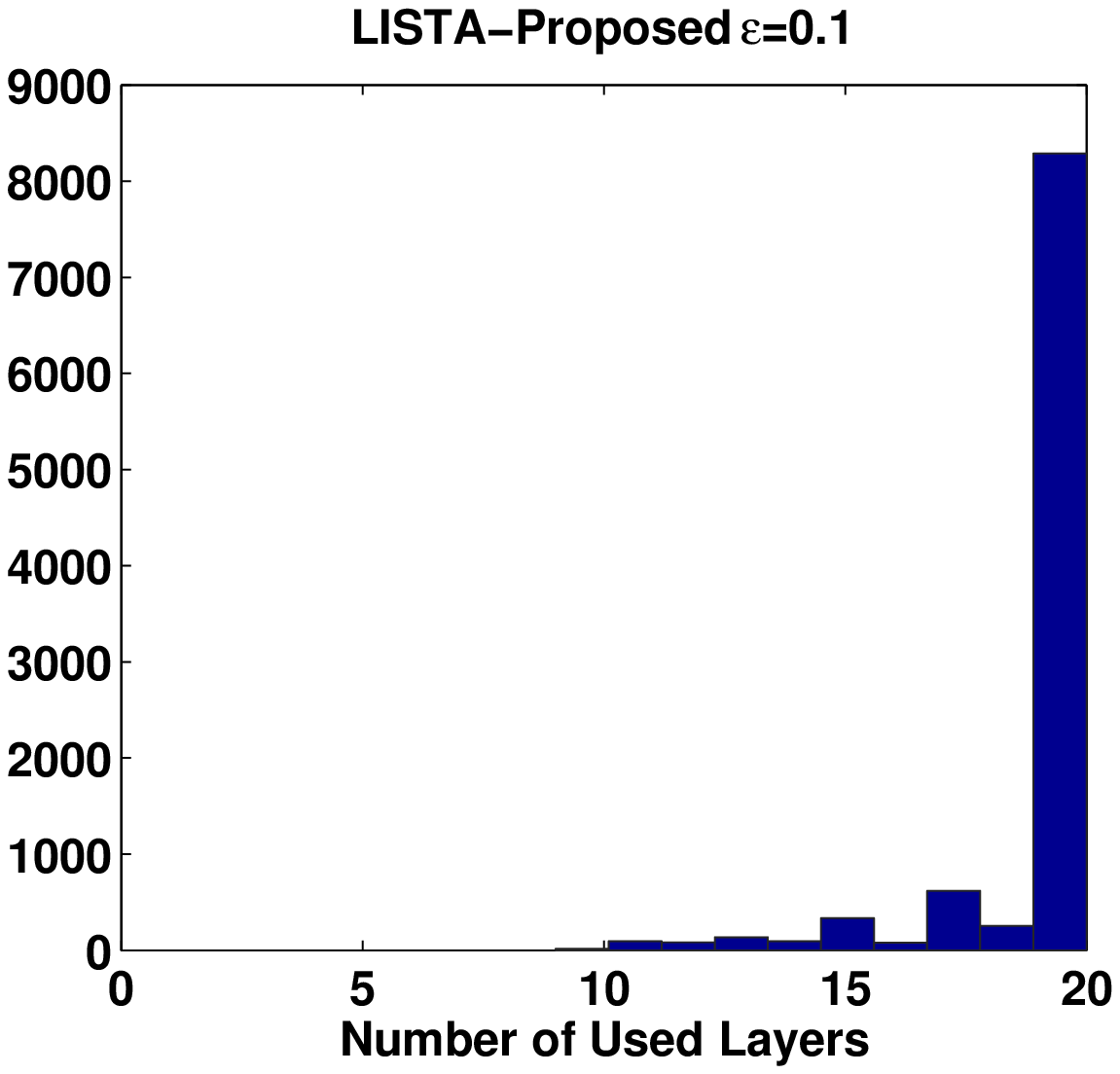}
\label{Dist-Layers1}}
\subfigure[Distribution of the number of used layers ($\epsilon=0.3$)]{\includegraphics[width=0.234\linewidth]{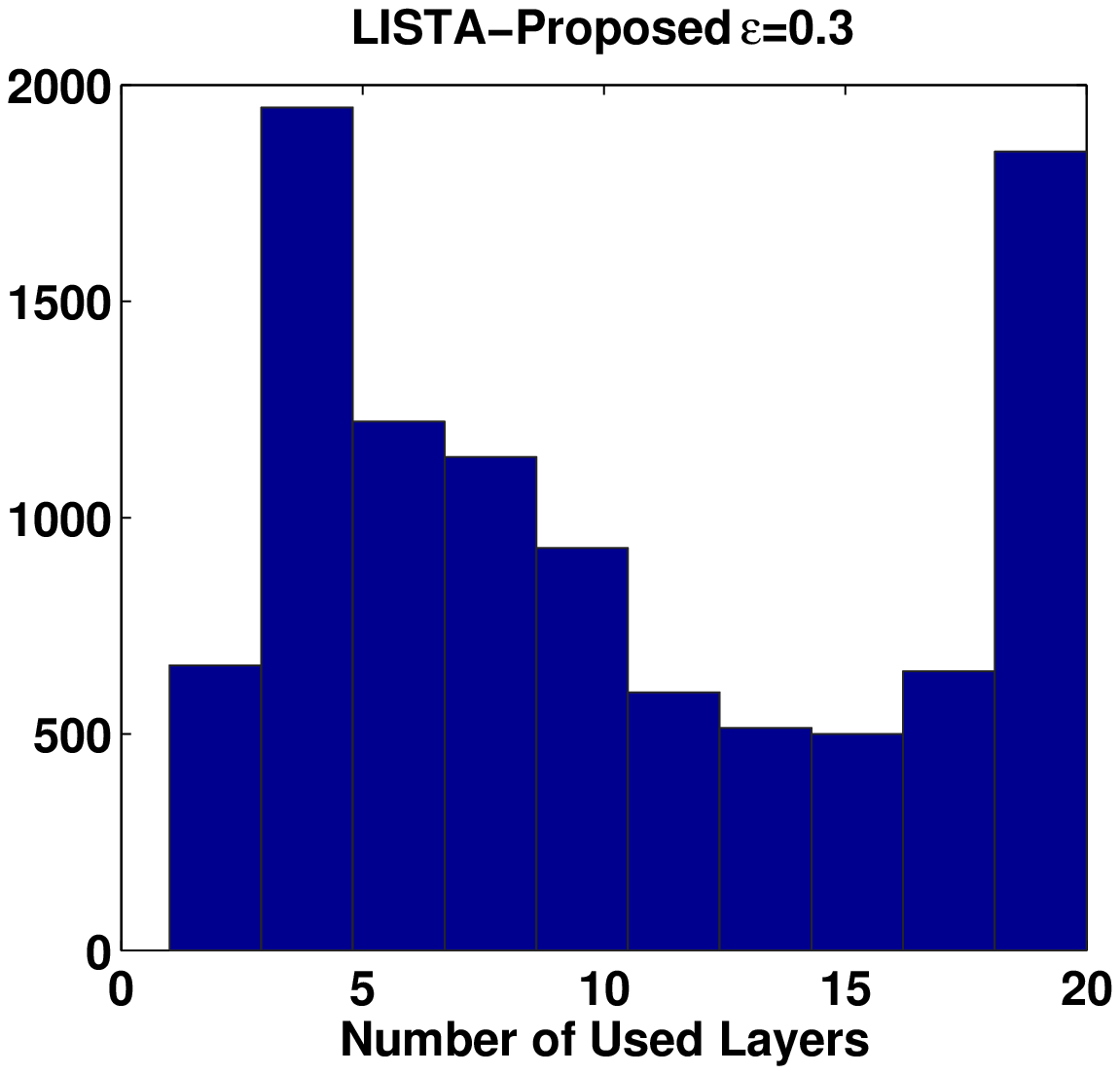}
\label{Dist-Layers3}}
\subfigure[Distribution of the number of used layers ($\epsilon=0.5$)]{\includegraphics[width=0.234\linewidth]{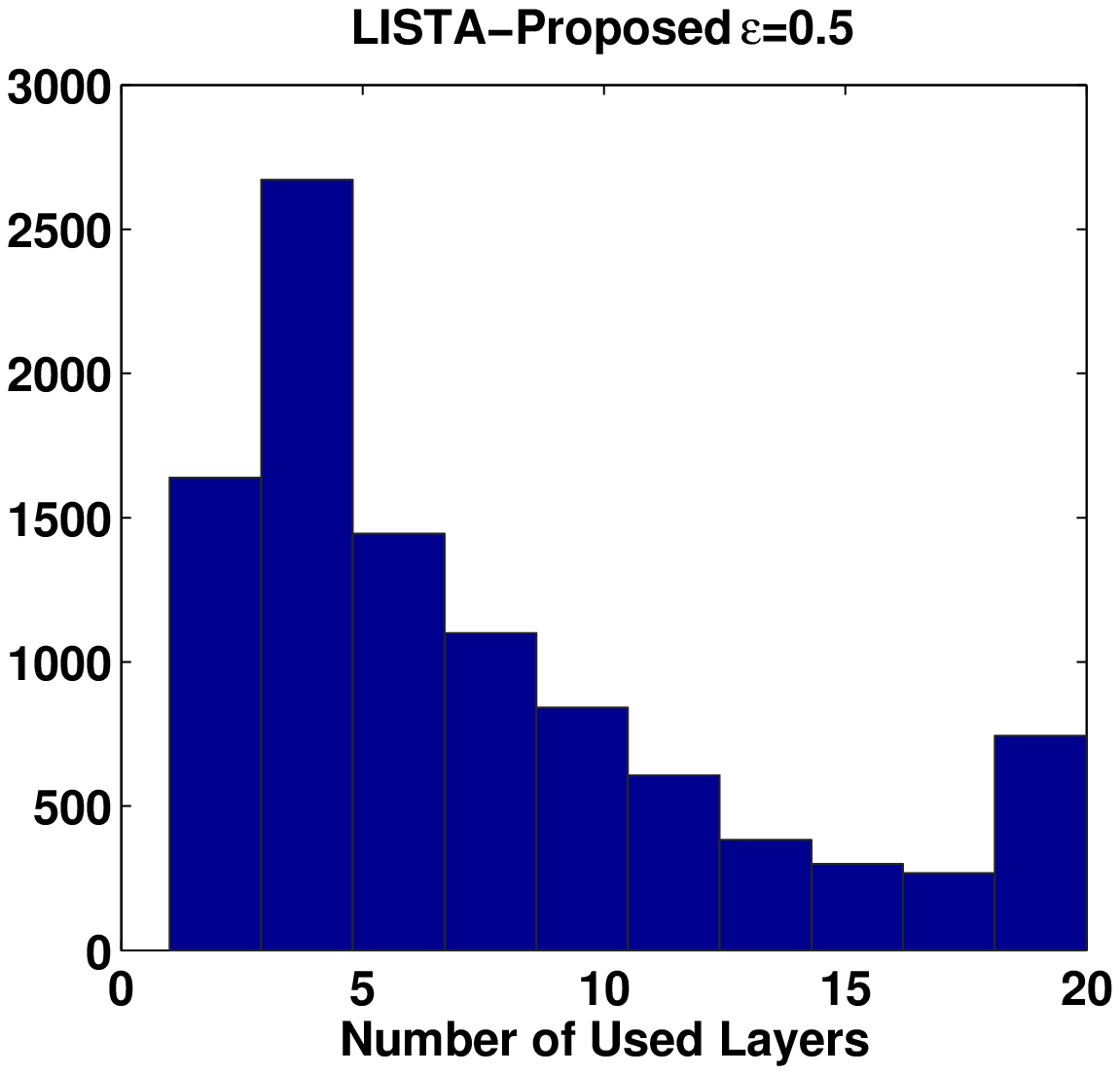}
\label{Dist-Layers5}}
\caption{Histogram of the sparsity level vs. histogram of the number of used layers of the proposed LISTA.} \label{fig:Dist}
\end{figure*}
\subsection{Compressive Random Access in Massive Machine-Type Communication}
Recent years observe a growing interest in massive MTC owing to the rapid development of Internet of Things and the 5th-generation (5G) wireless communications. In a typical MTC communication scene, a massive number of nodes sporadically transmit small packets with a low data rate, which is quite different to current cellular systems that are designed to support high data rates and reliable connections of a small number of users per cell. Communication overhead takes up a larger portion of resources in the MTC scene, and thus more efficient access methods are needed.

\par
One potential approach to reduce the communication overhead is the compressive sensing (CS) based multiuser detection (MUD)~\cite{bockelmann2013compressive}, which reduces communication overhead by eliminating control signaling. In CS-MUD, each user is assigned a unique pilot sequence, which is transmitted when the user needs to access the base station (BS). Then the random access problem can be formulated as the sparse linear inverse problem in (\ref{eq:sparse_problem}), where columns of $\mathbf{A}$ denote pilot sequences of different users, $\mathbf{y}$ denotes the signal received by the BS, and elements in $\mathbf{x}$ denote user activity and channel information. $x_j=0$ means the $j$th user is inactive.

\par
The number of active users is difficult to predict, and could vary in a large range depending on service type and user mobility. If few users are active for access, using a fixed number of iterations or depth would lead to the waste of computing power and increase communication latency (especially when the number of nodes is large). If a relatively large number of users are active, using a fixed number of iterations or depth would lead to poor user detection accuracy, as the algorithm is not likely to converge yet. Therefore, the proposed method for adaptive depth provides the solution to make the neural network adapt to the varying condition in massive MTC systems.

\par
In our numerical experiments, the pilots in $\mathbf{A}$ are randomly generated as i.i.d. QPSK, i.e., $\{\pm1,\pm i\}$ where $i=-\sqrt{1}$. We
assumed $m = 256$ users, pilots of length $n = 64$, and additive receiver noise of 20 dB. In the training, each mini-batch include 1024 random draws of $\mathbf{x}$ with sparsity level in the range $[1,20]$. The regularization parameter is set as $\tau=100$. For the training, the learning rate adjustment mechanism is the same as the previous synthetic experiments. For the inference, we used the same $\mathbf{A}$, and each reported result is averaged over 10000 mini-batches. We consider to use the LISTA, the LISTA-CPSS and the LAMP, i.e., DL based algorithms, for the CS-MUD that involves solving a linear inverse problem. The depth of the traditional networks is set as 18 while the maximum depth of proposed networks is set as 20. By varying the value of the halting constant $\varepsilon$ in (\ref{eq:halting_score_prop}), the proposed networks ejects outputs at different layers, and we evaluate the reconstruction performance with different averaged numbers of executing layers. As shown Fig.~\ref{fig:random-access}, the proposed approach leads to the decrease of reconstruction error for the LISTA, the LAMP and the LISTA-CPSS. In addition, the distributions of the sparsity level of user activity and the number of used layers of the proposed LISTA are shown in Fig.~\ref{fig:Dist} (a) and Fig.~\ref{fig:Dist} (b)-(d), respectively. It demonstrates the adaptive computing time characteristics of the proposed method for different sparsity levels. The results in the figure look as expected: more layers are executed for a smaller halting constant $\varepsilon$.

\subsection{Massive MIMO Channel Estimation}
Massive MIMO is another technology developed for 5G wireless communication systems, where the BS is equipped with a very large number of antennas to improve spectral and energy efficiency. In the time division duplex (TDD) transmission mode, the channel estimation is based on channel reciprocity, and the estimated channel in the uplink can be used for precoding in the downlink. According to the 5G channel model in~\cite{7934066}, the per-angle channel coefficients are sparse, as few users contribute significant energy to a given receive angle. Furthermore, owing to the grouped scatterers, individual multipath components with varying delays, angle-of-arrivals and angle-of-departures form several clusters~\cite{6375940}. Then the channel estimation can be casted as multiple sparse linear inverse problems corresponding to different angle-of-arrivals and angle-of-departures.

\par
\begin{figure}[!tb]%
\centering%
\includegraphics[width=0.45\textwidth]{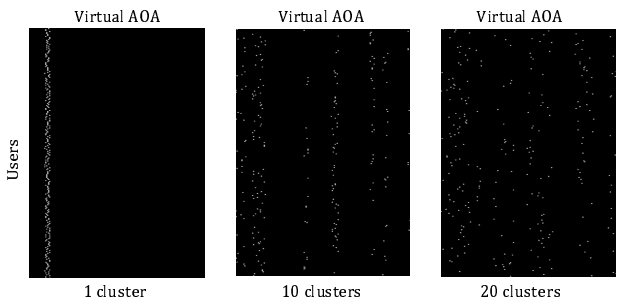}%
\DeclareGraphicsExtensions. \caption{Illustration of the massive MIMO channel with different number of clusters. } \label{fig:MIMO-cluster}
\end{figure}

In this experiment, we assumed i.i.d. QPSK pilots and one primary hexagonal cell with $m = 256$ users within the cell. Each user is associated with one pilot sequence of length $n = 128$, which is one column from $\mathbf{A}$. The BS is equipped with $64$ antennas, and the receiver noise is AWGN with the SNR of 20 dB. We generate the channel using the code provided by the authors in~\cite{7934066} (more details given in Appendix B of~\cite{7934066}). For simplicity, we set the users that have similar angle-of-arrivals as one cluster, and set the number of clusters uniformly distributed in $[1,20]$. In each cluster, the angular spread of angle-of-arrivals is $10^{\circ}$. The depth of traditional methods is set as 4 while the maximum depth of the proposed method is set as 6. Fig~\ref{fig:MIMO-cluster} illustrates the angle-of-arrivals of users with different number of clusters. Since users in one cluster contribute to few arrival directions, after transforming the multi-antennas' channel measurement matrix into the angular domain, each column of the transformed matrix corresponds to one sparse linear inverse problem. Obviously,we have different sparsity levels when the number of clusters varies.

\begin{figure}[!tb]%
\centering%
\includegraphics[width=0.45\textwidth]{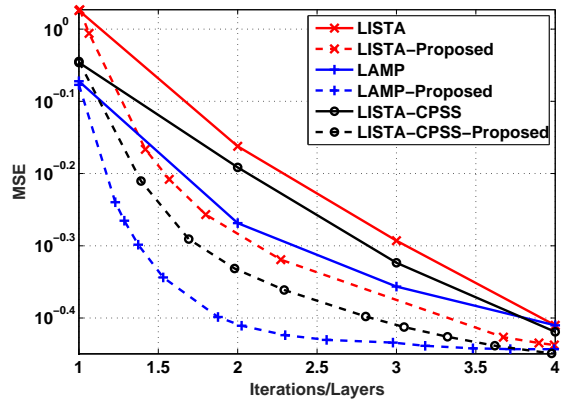}%
\DeclareGraphicsExtensions. \caption{Comparison of the reconstruction error with different averaged numbers of executing layers for massive MIMO channel estimation. } \label{fig:MIMO}
\end{figure}%

\par
The result of reconstruction error versus averaged numbers of executing layers for massive MIMO channel estimation is reported in Fig.~\ref{fig:MIMO}. Again, improved performance can be observed in extending any traditional method with the proposed method. For example, to achieve the MSE of $10^{-0.3}$, the LISTA requires about 3 layers in average, while the proposed method only needs about 2 layers, which means a reduction of 30 percent depth without sacrificing estimation accuracy.

\section{Conclusion}
This paper introduce a DL based method with adaptive network depth for solving sparse linear inverse problem with applications in wireless communications. The proposed method is able to automatically adjust the number of iterations/layers for different tasks with varying degree of hardness. In comparison with existing methods in literature, the innovations of the proposed network include the new design for the halting score, the construction of network output, the cost function and the nonsymmetric training-testing process. Theoretical convergence analysis for the PGD algorithm in two different cases, i.e., PGD with a learned gradient and PGD with adaptive depth, is provided, which sheds lights on the inside of the adaptive depth mechanism in solving sparse linear inverse problem. Experiment using both synthetic data and applications including random access in massive MTC and massive MIMO channel estimation demonstrate the improved efficiency for the proposed approach.


\appendices

\section{Parameter Update Steps for the Proposed Cost Function (\ref{eq:cost_function_1})}
We first consider the updates of the parameters in the halting network. According to the cost function (\ref{eq:cost_function_1}), the partial derivative regarding to $h_{t}$ can be derived as
\begin{equation}
\begin{split}
\frac{\partial \mathcal{L}}{\partial h_{t}}=\tau- \frac{\|\mathbf{x}-\mathbf{x}_{t}\|_{2}^{2}}{h_{t}^{2}}.
\end{split}
\end{equation}
The derivatives of halting score function (\ref{eq:halting_score}) can be expressed as
\begin{equation}
\begin{split}
\frac{\partial h_{t}}{\partial \phi_{t}}=&\sigma(\phi_{t}\|\mathbf{Q}(\mathbf{y}-\mathbf{Ax}_{t})\|^{2}_{2}+\varphi_{t})\\
&(1-\sigma(\phi_{t}\|\mathbf{Q}(\mathbf{y}-\mathbf{Ax}_{t})\|^{2}_{2}+\varphi_{t}))
\|\mathbf{Q}(\mathbf{y}-\mathbf{Ax}_{t})\|^{2}_{2},
\end{split}
\end{equation}

\begin{equation}
\begin{split}
\frac{\partial h_{t}}{\partial \varphi_{t}}=&\sigma(\phi_{t}\|\mathbf{Q}(\mathbf{y}-\mathbf{Ax}_{t})\|^{2}_{2}+\varphi_{t})\\
&(1-\sigma(\phi_{t}\|\mathbf{Q}(\mathbf{y}-\mathbf{Ax}_{t})\|^{2}_{2}+\varphi_{t})),
\end{split}
\end{equation}

\begin{equation}
\begin{split}
\frac{\partial h_{t}}{\partial \mathbf{Q}}=&\sigma(\phi_{t}\|\mathbf{Q}(\mathbf{y}-\mathbf{Ax}_{t})\|^{2}_{2}+\varphi_{t})\\
&(1-\sigma(\phi_{t}\|\mathbf{Q}(\mathbf{y}-\mathbf{Ax}_{t})\|^{2}_{2}+\varphi_{t}))\\
&2\phi_{t}\mathbf{Q}(\mathbf{y}-\mathbf{Ax}_{t})(\mathbf{y}-\mathbf{Ax}_{t})^{T},
\end{split}
\end{equation}
\begin{equation}
\begin{split}
\frac{\partial h_{t}}{\partial \mathbf{x}_{t}}=&\sigma(\phi_{t}\|\mathbf{Q}(\mathbf{y}-\mathbf{Ax}_{t})\|^{2}_{2}+\varphi_{t})\\
&(1-\sigma(\phi_{t}\|\mathbf{Q}(\mathbf{y}-\mathbf{Ax}_{t})\|^{2}_{2}+\varphi_{t}))\\
&2\phi_{t}\mathbf{A}^{T}\mathbf{Q}^{T}\mathbf{Q}(\mathbf{y}-\mathbf{Ax}_{t}).
\end{split}
\end{equation}

Given the above derivatives, the updates of the parameters in the halting network are given as following
\begin{equation}
\begin{split}
\frac{\partial \mathcal{L}}{\partial \phi_{t}}=\frac{\partial \mathcal{L}}{\partial h_{t}}\frac{\partial h_{t}}{\partial \phi_{t}},\\
\frac{\partial \mathcal{L}}{\partial \varphi_{t}}=\frac{\partial \mathcal{L}}{\partial h_{t}}\frac{\partial h_{t}}{\partial \varphi_{t}},\\
\frac{\partial \mathcal{L}}{\partial \mathbf{Q}}=\sum_{t=1}^{L}\frac{\partial \mathcal{L}}{\partial h_{t}}\frac{\partial h_{t}}{\partial \mathbf{Q}}.
\end{split}
\end{equation}

\par
Now we consider the update of the parameters in the deep unfolding networks. To simplify notification, we define $l(\mathbf{x}_{t})=\frac{\|\mathbf{x}-\mathbf{x}_{t}\|_{2}^{2}}{h_{t}}+\tau h_{t}$, where $l(\mathbf{x}_{t})$ is a function of $\mathbf{x}_{t}$. Then the cost function (\ref{eq:cost_function_1}) can be rewritten as
\begin{equation}
\begin{split}
\mathcal{L}(\boldsymbol{\theta})=\sum_{t=1}^{L}l(\mathbf{x}_{t}).
\end{split}
\end{equation}
The derivative of $l(\mathbf{x}_{t})$ is
\begin{equation}
\begin{split}
\frac{\partial l(\mathbf{x}_{t})}{\partial \mathbf{x}_{t}}=\frac{-2(\mathbf{x}-\mathbf{x}_{t})}{h_{t}}+\left(\tau-\frac{\|\mathbf{x}-\mathbf{x}_{t}\|_{2}^{2}}{h_{t}^{2}}\right)\frac{\partial h_{t}}{\partial \mathbf{x}_{t}}\approx\mathcal{O}\left(\frac{1}{h_{t}^{2}}\right)
\end{split}
\end{equation}

\par
For the last layer, $h_{L}$ is a predefined small constant to assure that all tasks can emit at the last layer $L$, so there would be no gradient flow from $h_{L}$, which leads to $\frac{\partial h_{L}}{\partial \mathbf{x}_{L}}=0$. Therefore, the partial derivative can be expressed as
\begin{equation}
\begin{split}
\frac{\partial \mathcal{L}}{\partial \mathbf{x}_{L}}=\frac{\partial l(\mathbf{x}_{L})}{\partial \mathbf{x}_{L}}=\frac{-2(\mathbf{x}-\mathbf{x}_{L})}{h_{L}}\approx\mathcal{O}(\frac{1}{h_{L}})
\end{split}
\end{equation}
For the layer $t=1,\ldots,L-1$, we have
\begin{equation}\label{eq:appendix_1}
\begin{split}
\frac{\partial \mathcal{L}}{\partial \mathbf{x}_{t}}&=\frac{\partial l(\mathbf{x}_{t})}{\partial \mathbf{x}_{t}}+\frac{\partial l(\mathbf{x}_{t+1})}{\partial \mathbf{x}_{t}}+
\cdots+\frac{\partial l(\mathbf{x}_{L})}{\partial \mathbf{x}_{t}}\\
&=\frac{\partial l(\mathbf{x}_{t})}{\partial \mathbf{x}_{t}}+(\frac{\partial \mathbf{x}_{t+1}}{\partial \mathbf{x}_{t}})^{T}\frac{\partial l(\mathbf{x}_{t+1})}{\partial \mathbf{x}_{t+1}}+
\cdots+(\frac{\partial \mathbf{x}_{L}}{\partial \mathbf{x}_{t}})^{T}\frac{\partial l(\mathbf{x}_{L})}{\partial \mathbf{x}_{L}}\\
&\approx \mathcal{O}(\frac{1}{h_{t}^{2}})+(\frac{\partial \mathbf{x}_{t+1}}{\partial \mathbf{x}_{t}})^{T}\mathcal{O}(\frac{1}{h_{t+1}^{2}})+
\cdots+(\frac{\partial \mathbf{x}_{L}}{\partial \mathbf{x}_{t}})^{T}\mathcal{O}(\frac{1}{h_{L}})
\end{split}
\end{equation}
Since $h_{L}$ is a predefined small constant satisfying $h_{L}\ll h_{t}^{2}$ ($t=1,\ldots,L-1$), the partial derivative in (\ref{eq:appendix_1}) can be approximated as
\begin{equation}\label{eq:appendix_2}
\begin{split}
\frac{\partial \mathcal{L}}{\partial \mathbf{x}_{t}}\approx (\frac{\partial \mathbf{x}_{L}}{\partial \mathbf{x}_{t}})^{T}\frac{\partial l(\mathbf{x}_{L})}{\partial \mathbf{x}_{L}}=(\frac{\partial \mathbf{x}_{L}}{\partial \mathbf{x}_{t}})^{T}\frac{-2(\mathbf{x}-\mathbf{x}_{L})}{h_{L}}
\end{split}
\end{equation}

\par
In the original deep unfolding networks (e.g., LISTA, LAMP and LISTA-CPSS) without the halting scheme, the loss function is $\mathcal{L}=\|\mathbf{x}-\mathbf{x}_{L}\|_{2}^{2}$. The partial derivative is
\begin{equation}\label{eq:appendix_3}
\begin{split}
 \frac{\partial \mathcal{L}}{\partial \mathbf{x}_{t}}=(\frac{\partial \mathbf{x}_{L}}{\partial \mathbf{x}_{t}})^{T}\frac{\partial \mathcal{L}}{\partial \mathbf{x}_{L}}=(\frac{\partial \mathbf{x}_{L}}{\partial \mathbf{x}_{t}})^{T}2(\mathbf{x}_{T}-\mathbf{x})
 \end{split}
\end{equation}
In comparison of (\ref{eq:appendix_2}) and (\ref{eq:appendix_3}), the partial derivatives for generating $x_{t}$ are same except for a fixed scaling factor $\frac{1}{h_{L}}$. Note that the fixed scaling factor can be offset by using a small learning rate in the training process. Therefore, the derivatives of the parameters in the deep unfolding network, e.g., $\mathbf{W}$ in the LISTA, are same in the two cases. To accelerate the learning of halting network, we propose a two-stage training process in Section V.




\bibliographystyle{IEEEbib}
\bibliography{bib_paper}

\end{document}